\documentclass{cernyrep}

\usepackage{varwidth}
\usepackage{xcolor}
\begin{document}
\title{Multidimensional Plasma Wake Excitation in the Non-linear Blowout Regime}
\author{J. Vieira$^1$, R.A. Fonseca$^{1,2}$ and L.O. Silva$^1$}
\institute{$^1$GoLP/Instituto de Plasmas e Fus\~{a}o Nuclear, Instituto Superior T\'{e}cnico, Universidade de Lisboa, Lisbon, Portugal \\
$^2$DCTI/ISCTE Lisbon University Institute, Lisbon, Portugal}
\maketitle

\begin{abstract}
Plasma accelerators can sustain very high acceleration gradients. They are promising candidates for future generations of particle accelerators for several scientific, medical and technological applications. Current plasma-based acceleration experiments operate in the relativistic regime, where the plasma response is strongly non-linear. We outline some of the key properties of wakefield excitation in these regimes. We outline a multidimensional theory for the excitation of plasma wakefields in connection with current experiments. We then use these results and provide design guidelines for the choice of laser and plasma parameters ensuring a stable laser wakefield accelerator that maximizes the quality of the accelerated electrons. We also discuss the role of structured laser pulse drivers to circumvent and expand the current limits of laser plasma accelerators.
\end{abstract}

\section{Introduction}

In their seminal work published more than 30 years ago~\cite{bib:tajima_prl_1979}, Toshiki Tajima and John Dawson proposed the concept of the laser wakefield accelerator. Through theoretical calculations and computer simulations they showed that the radiation pressure of an intense laser pulse could drive large-amplitude plasma waves with a phase velocity identical to the group velocity of the laser driver, and characterized by large accelerating electric fields. These electric fields could then accelerate electrons to high energies in very short distances when compared with conventional accelerating devices. This work initiated research efforts that have also motivated the construction of some of the most powerful lasers available today~\cite{bib:eli}.

Laser wakefield accelerators can be thought as energy transformers, converting the energy from the laser driver to the energy of plasma waves, and then converting the energy of the plasma waves to the accelerated particles. The laser acts on the plasma through the ponderomotive force, which expels particles from the regions where the laser fields are more intense to the regions of lower intensities. This force can also be interpreted as the radiation pressure that an intense laser exerts on background plasma electrons~\cite{bib:silva_pre_1999}. The displacement of background plasma electrons leads to large electrostatic fields due to the space--charge separation between the background plasma electrons and ions, which remain immobile. Typically, these plasma electrostatic fields scale with $E_0 \mathrm{[V cm^{-3}]}\simeq 0.96 \sqrt{n_0~[\mathrm{cm^{-3}}]}$, where $n_0$ is the background plasma density. Thus, typical plasmas densities of $n_0 \simeq 10^{18}~\mathrm{cm}^{-3}$ can lead to accelerating electric fields in excess of 1~GV cm$^{-1}$~\cite{bib:esarey_ieee_1996}.

In order to excite these very high accelerating fields, current experiments typically use tightly focused, high-intensity and ultra-short laser pulses with transverse spot sizes smaller than $100~\mathrm{\mu m}$, intensities above $I\sim 10^{18}$ W cm$^{-2}$ and pulse durations shorter than $100~\mathrm{fs}$. Lasers with these properties were not available when the laser wakefield accelerator was first proposed. Recently, however, with the advances in laser technologies, these lasers started to become widely available for experiments. The first successful experiments capable of generating electron bunches with non-Maxwellian energy distributions were independently reported in 2004 by three experimental groups. These experiments~\cite{bib:geddes_nature_2004,bib:faure_nature_2004,bib:mangles_nature_2004} used ${\sim}1$ J lasers, focused to transverse spot sizes of ${\sim}10~\mathrm{\mu m}$ and compressed down to $30~\mathrm{fs}$. The lasers hit ${\sim}1~\mathrm{mm}$ long gas jets producing plasmas with densities on the order of $10^{19}~\mathrm{cm}^{-3}$. After the gas jet, the experiments measured a ${\sim}10\%$ energy spread for 50--120~$\mathrm{MeV}$ electron bunches at distances ranging between 1 and 3~$\mathrm{mm}$.

These experimental progresses took place even without the advanced theoretical and conceptual plasma-based acceleration framework that is currently available. The first theoretical and computational results for plasma wave excitation and electron acceleration were obtained in the one-dimensional (1D) limit, because the equations can be solved exactly in this limit. The 1D limit, however, presents fundamental limitations that are inherent to the reduced dimensionality. For instance, the motion of free plasma electron oscillations in one dimension is well described by an harmonic oscillator. In multidimensions, however, the plasma single electron trajectories are described by an anharmonic oscillator even for low-amplitude plasma waves~\cite{bib:dawson_pr_1959}. As a result, the period of the oscillations depends on their amplitude of oscillation. Thus, in one dimension, if the background plasma electron flow is laminar during the first plasma oscillation it will remain laminar. On the contrary, the electron flow will inevitably become turbulent in multidimensions~\cite{bib:dawson_pr_1959}.

The electromagnetic structure of the plasma wakefield depends mainly on the amplitude of the plasma electron oscillations. Current plasma-based acceleration experiments operate in strongly non-linear regimes, for which the amplitude of the plasma electron radial displacement is much larger than its initial radial position. In this regime, plasma electron trajectories become non-laminar before the end of the first plasma oscillation leading to plasma wave-breaking~\cite{bib:leemans_prl_2014,bib:litos_nature_2014}. Although simplified analytical models exist to describe the structure of the wakefield, the full electromagnetic field structure of the plasma wave can only be captured through numerical simulations. Simulations have then been used to plan and to predict the experimental results, which is also essential to develop and confirm the predictions of the analytical models.

There are several numerical models that can describe plasma acceleration computationally, each employing different approximations. One of the most successful techniques is the particle-in-cell (PIC) method~\cite{bib:osiris,bib:quickpic}. In PIC codes, space is discretized into a grid that stores electric and magnetic fields. Each grid cell contains simulation macro-particles, each representing an ensemble of real electrons. It is then possible to advance the particles positions and momenta using the Lorentz force equation by interpolating the fields at the particles positions. Current densities, which are defined at the edges of the grid, are used to advance the fields through a set of discretized Maxwell's equations. As a result, PIC modelling retains the kinetic nature of the plasma dynamics, and can be employed to describe the plasma even in turbulent regimes where the flow of the plasma electrons is non-laminar.

There are several categories of PIC codes depending on the physics they retain/neglect. Reduced PIC codes use reduced versions of the Lorentz force (e.g., by neglecting relativistic effects), or reduced sets of Maxwell's equations (e.g., by neglecting magnetic fields). In spite of not retaining the full physics, these codes are typically very computationally effective, and allow for fast turn around simulation times when compared with full PIC codes. Full PIC codes employ almost no physical approximations. Particles advance under the relativistic Lorentz force, and the fields are updated with the full set of Maxwell's equations. Full PIC simulations are well suited to describe plasma accelerators in strongly non-linear regimes, but they are very computationally expensive, requiring the use of large super-computers. In order to optimize available computational infrastructures, and in order to better assist experimental design and interpretation, PIC algorithms have also seen significant advances. The most efficient PIC algorithms can currently achieve very high computing efficiencies from a few thousands up to more than a million cpu-cores~\cite{bib:fonseca_ppcf_2013}. One of the largest laser wakefield acceleration (LWFA) simulations sucessfuly ran in more than 200,000 cores, and enabled the simulation of fine details regarding the acceleration process, while reaching petascale sustainable performance in a production run~\cite{bib:fonseca_ppcf_2013}.

Here we will review some of the key advances regarding wake excitation in the linear and the non-linear regimes, in one dimension and in multidimensions, complementing the analytical results with PIC simulation results. In Section~\ref{sec:nonlinear1d} we will describe a theory for linear and non-linear plasma wave excitation in one-dimension, and also discuss the physics of beam loading in one dimension. In Section~\ref{sec:fenoblowout} we will derive a phenomenological theory for multidimensional, non-linear plasma waves, capable of describing current experiments. We will then derive an analytical theory for the blowout regime in Section~\ref{sec:blowout} that predicts the full electromagnetic structure of the plasma wave. In Section~\ref{sec:beamloading} we will use the theoretical framework derived in Section~\ref{sec:blowout} in order to explore the physics of beam loading in the blowout regime. We also discuss the role of structured laser pulses to expand plasma accelerators beyond current possibilities in Sec.~\ref{sec:structuredlight} and in Section~\ref{sec:conclusions} we present the conclusions.

\section{Non-linear plasma waves in the 1D limit}
\label{sec:nonlinear1d}

In order to introduce the key physical mechanisms of plasma acceleration, we will first explore the self-consistent generation of plasma waves by a laser pulse in non-weakly relativistic or linear regimes and in the 1D limit.

\subsection{Relativistic fluid and Maxwell's equations in weakly relativistic regimes}
\label{subsec:linear1d}

In order to explore the dynamics in one dimension we consider the limit of wide laser pulses, with transverse spot-sizes much wider than the plasma wavelength. In addition, we will also consider the excitation of linear plasma waves characterized by small density perturbations when compared to the background plasma density. This excitation regime of the plasma waves, usually characterized by sinusoidal plasma density and electric field oscillations, corresponds to the so-called weakly relativistic or linear regime. Our starting point are Maxwell's equations written in the Coulomb gauge for the laser vector potential $\mathbf{A}$, which read
\begin{equation}
\label{eq:waveeq_full}
\frac{1}{c^2}\frac{\partial^2 \mathbf{A}}{\partial t^2} + \nabla\times\nabla\times\mathbf{A}=\frac{4 \pi}{c}\mathbf{J}-\frac{1}{c}\frac{\partial \nabla \phi}{\partial t},
\end{equation}
where $t$ is the time, $\mathbf{J}=n \mathbf{v}$ is the total electric current, $\phi$ is the scalar (electrostatic) potential and $c$ the speed of light. In addition, $n$ is the local plasma density and $\mathbf{v}$ is the local plasma fluid velocity.

We can rewrite Eq.~(\ref{eq:waveeq_full}) in the direction of polarisation of a linearly polarized laser pulse. We then assume that the laser is polarized in $y$ and propagates in $x$. The source term for Eq.~(\ref{eq:waveeq_full}) is the transverse plasma current driven by the laser fields, $J_y$. In order to determine $J_y$, we assume that the plasma ions remain fixed during the interaction. This is a valid approximation because the mass of background plasma ions ($m_\mathrm{i}$) is much higher than the electron mass $m_\mathrm{e}$. Thus, plasma ions barely move during a plasma electron oscillation. The plasma currents are then only due to the motion of plasma electrons. Conservation of canonical momentum for the plasma electrons in the $y$ direction, assumed to be initially at rest, implies that $p_y = e A_y/(m_\mathrm{e} c^2)$, where $p_y$ is the electron momentum in the $y$ direction, and where $e$ and $m_\mathrm{e}$ are the electron charge and mass, respectively. Since $p_y = v_y \gamma$, the transverse plasma current driven by the laser is simply given by $J_y = n v_y = n p_y/\gamma = n A_y /(m_\mathrm{e} c \gamma)$. In the former expression, $\gamma$ is the relativistic factor and $v_y$ is the electron velocity in the $y$ direction. Using this expression for $J_y$, Eq.~(\ref{eq:waveeq_full}) then becomes
\begin{equation}
\label{eq:waveeq_1d}
\frac{1}{c^2}\frac{\partial^2 A_y}{\partial t^2} + \frac{\partial^2 A_y}{\partial x^2} =\frac{4 \pi}{c}J_y -\frac{1}{c}\frac{\partial \phi}{\partial t} = - \frac{4 \pi e^2}{m_\mathrm{e} c^2}\frac{n}{\gamma} A_y,
\end{equation}
where we have used the fact that in one dimension $\partial \phi/\partial y = 0$. We note that the conservation of canonical momentum is strictly valid for the case of plane electromagnetic waves. Thus, although valid exactly in 1D geometries, the conservation of canonical momentum does not hold exactly in multidimensions when the laser pulse has a finite transverse spot-size.

In order to further evaluate Eq.~(\ref{eq:waveeq_1d}), we need to find an expression for the relativistic $\gamma$ factor involving only known quantities. The relativistic factor $\gamma$ of plasma electrons is given by
\begin{equation}
\label{eq:gamma}
\gamma = \sqrt{1 + \frac{p_x^2}{m_\mathrm{e}^2 c^2} + \frac{p_y^2}{m_\mathrm{e}^2 c^2} + \frac{p_z^2}{m_\mathrm{e}^2 c^2} },
\end{equation}
where $p_x$ is the longitudinal electron momentum (in the $x$ direction) and where $p_z$ is the transverse electron momentum in $z$. Although plasma electrons can have transverse velocities in $y$ and in $z$ (as long as the laser has electric field components in $y$ or in $z$), the motion of plasma electrons is only along $x$ in one-dimension. Since we have assumed that the laser is polarized in $y$, only $v_y\ne 0$.

As with Eq.~(\ref{eq:waveeq_1d}), we can rewrite the relativistic factor given by Eq.~(\ref{eq:gamma}) using conservation of canonical momentum in $y$. Hence, since the laser is polarized in $y$, $p_y/(m_\mathrm{e} c) = e A_y/(m_\mathrm{e} c^2)$ and $p_z = 0$. Here it is worth noting that $p_y/(m_\mathrm{e} c) = \gamma v_y$ is the proper electron velocity and that the quantity $e A_y/ (m_\mathrm{e} c^2)=a_y$ is the normalized vector potential. Thus, the conservation of canonical momentum indicates that the normalized laser vector potential, $a_y$, can be regarded as the momentum associated with the fast quiver motion of electrons on the laser fields. Therefore, the motion of the plasma electrons becomes relativistic when $p_y/(m_\mathrm{e} c) = a_y \gtrsim 1$, and remains non- or weakly-relativistic for $a_y \ll 1$. Equation~(\ref{eq:gamma}) can then be simplified by expanding the relativistic factor for small $p_x/m_\mathrm{e} c\ll 1$ and small $a_y\ll 1$, yielding
\begin{equation}
\label{eq:gammaapprox}
\gamma = \sqrt{1 + \frac{p_x^2}{m_\mathrm{e}^2 c^2} + \frac{e^2 A_y^2}{m_\mathrm{e}^2 c^4}} \simeq 1 + \frac{1}{2} \frac{p_x^2}{m_\mathrm{e}^2 c^2} + \frac{1}{2} \frac{e^2 A_y^2}{m_\mathrm{e}^2 c^4}.
\end{equation}
We can hence write the $1/\gamma$ factor appearing in the wave equation Eq.~(\ref{eq:waveeq_1d}) as
\begin{equation}
\label{eq:gammainv}
\frac{1}{\gamma} \simeq 1 - \frac{1}{2} \frac{e^2 A_y^2}{m_\mathrm{e}^2 c^4} - \frac{1}{2} \frac{p_x^2}{m_\mathrm{e}^2 c^2},
\end{equation}
which after being inserted into the simplied wave Eq.~(\ref{eq:waveeq_1d}) gives
\begin{equation}
\label{eq:waveeq_plasma}
\frac{1}{c^2}\frac{\partial^2 A_y}{\partial t^2} + \frac{\partial^2 A_y}{\partial x^2} \simeq - \frac{4 \pi e^2}{m_\mathrm{e} c^2} n \left(1 - \frac{1}{2} \frac{p_x^2}{m_\mathrm{e}^2 c^2} - \frac{1}{2} \frac{e^2 A_y^2}{m_\mathrm{e}^2 c^4}\right) A_y.
\end{equation}
We can further simplify Eq.~(\ref{eq:waveeq_plasma}) by using the following ordering to obtain the 1D laser wave equation valid in non/weakly relativistic regimes, which we will confirm \emph{a posteriori}:
\begin{equation}
\label{eq:ordering}
\mathcal{O}\left(\frac{p_x}{m_\mathrm{e} c}\right) \simeq \mathcal{O}\left(\frac{e^2 A_y^2}{m_\mathrm{e}^2 c^4}\right) \simeq \mathcal{O}\left(1-\frac{\delta n}{n_0}\right)\ll 1,
\end{equation}
where the plasma electron density perturbations are defined according to $n = n_0 + \delta n$, where $n_0$ is the background plasma density and $\delta n$ is a small perturbation. As a result, the wave equation Eq.~(\ref{eq:ordering}) reduces to
\begin{equation}
\label{eq:waveeq_plasma_2}
\frac{1}{c^2}\frac{\partial^2 A_y}{\partial t^2} + \frac{\partial^2 A_y}{\partial x^2} \simeq -\frac{\omega_{p0}^2}{c^2}\left(1+\frac{\delta n}{n_0}-\frac{1}{2}\frac{e^2 A_y^2}{m_\mathrm{e}^2 c^4}\right)A_y.
\end{equation}

Equation~(\ref{eq:waveeq_plasma_2}) has two unknowns, $A_y$ and $\delta n$. In order to close the model we now need an equation for $\delta n$. In order to determine an expression for $\delta n$, we consider the linearized continuity equation for the plasma electrons and its time derivative, which read
\begin{equation}
\label{eq:fluid}
\frac{\partial \delta n}{\partial t} + n_0 \nabla \cdot \left( \delta n \mathbf{v} \right) = 0 \Rightarrow \frac{\partial^2 \delta n}{\partial t^2} + n_0 \nabla \cdot \frac{\partial \mathbf{v}}{\partial t} = 0,
\end{equation}
where $\mathbf{v}\ll c $ is given by the Lorentz force equation for non/weakly relativistic regimes
\begin{equation}
\label{eq:force_1d}
\frac{\partial \mathbf{v}}{\partial t} = -\frac{e}{m_\mathrm{e}} \mathbf{E} - c^2\nabla\left(1+\frac{1}{2}\frac{e^2 A_y^2}{m_\mathrm{e}^2 c^4}\right),
\end{equation}
where $\mathbf{E}$ is the electric field and where the second term on the right-hand side of Eq.~(\ref{eq:force_1d}) is the laser ponderomotive force. Equation~(\ref{eq:force_1d}) neglects the magnetic field force component, a valid assumption because $\mathbf{v}\ll c$. Note that we have assumed that $\gamma = 1$ in Eq.~(\ref{eq:force_1d}). Inclusion of the higher order terms of Eq.~(\ref{eq:gammaapprox}) for the relativistic factor of plasma electrons in Eq.~(\ref{eq:force_1d}) would also lead to higher order relativistic corrections to the Lorentz force. Taking the divergence of Eq.~(\ref{eq:force_1d}), inserting the resulting expression into the continuity equation and using Gauss's law $\nabla \cdot E \simeq \delta n$ then gives
\begin{equation}
\label{eq:plasmawave}
\left(\frac{\partial^2}{\partial t^2} + \omega_\mathrm{p}^2\right)\frac{\delta n}{n_0} = \frac{1}{2}\frac{e^2}{m_\mathrm{e}^2 c^2} \ \nabla A_y^2.
\end{equation}
Equation~(\ref{eq:plasmawave}) closes the 1D model for the self-consistent laser-plasma interaction valid in non/weakly relativistic regimes and in one dimension. We note that Eq.~(\ref{eq:plasmawave}) confirms the ordering given by Eq.~(\ref{eq:ordering}) in that $\mathcal{O}\left(\delta n/n_0\right) \simeq \mathcal{O}\left(A_y^2\right)$. Equation~(\ref{eq:plasmawave}) is a forced harmonic oscillator excited by the laser radiation pressure (ponderomotive force). The laser expels electrons from regions of maximum laser intensity to regions of lower laser intensity. Background plasma ions provide a restoring force that attracts plasma electrons back to their initial positions. The generation of plasma waves in one dimension then occurs as follows: the laser starts by pushing plasma electrons forward. Since the ions remain fixed, a space charge electrostatic field develops, pushing the plasma electrons back to their original position. When they return to their original longitudinal position, the plasma electrons have a longitudinal backward velocity. They will then continue to move backwards with respect to their initial positions. A space charge force due to the background plasma ions forms again, pulling the plasma electrons forwards. This forms a plasma oscillation. The natural frequency of oscillation is the plasma frequency $\omega_{p}$.

\subsection{1D wakefield excitation by ultra-short lasers in non-linear regimes}

After having introduced the excitation of plasma waves in weakly relativistic regimes, we now generalize the 1D model given by Eqs.~(\ref{eq:waveeq_plasma_2}) and (\ref{eq:plasmawave}) to the relativistic and strongly non-linear regimes in the limit of short lasers compared with the plasma period.

It is useful to include dimensionless quantities. Plasma electric fields are then normalized to the cold wave-breaking limit ($E_{\mathrm{wb}}$). The cold wave-breaking limit corresponds to the maximum amplitude that a plasma wave supports in 1D limit and in the non-relativistic regime. Magnetic fields are normalized to the cold wave-breaking limit multiplied by the speed of light $c$ ($B_0$), scalar and vector potentials normalized to the electron rest energy divided by the elementary charge ($\phi_0$), and space and time are normalized to the plasma skin depth ($d_0$) and inverse plasma frequency $t_0$, respectively. In addition, momentum is normalized to $m_\mathrm{e} c$ and energy to $m_\mathrm{e} c^2$. In practical units, dimensionless quantities are then expressed as
\begin{eqnarray}
 \label{eq:norm}
 E_{\mathrm{wb}} & = &\frac{m_\mathrm{e} c \omega_\mathrm{p} }{e} \simeq 5.64\times10^4\sqrt{n_0\mathrm{[cm^{-2}]}}\,\mathrm{V/cm}, \\
 B_{0} & = & \frac{m_\mathrm{e} c^2 \omega_\mathrm{p} }{e} \simeq 32 \sqrt{n_0 \mathrm{[\times10^{16}~cm^{-3}]}}\,\mathrm{T,}\\
 \phi_0 \simeq \mathbf{A}_0 & = & \frac{m_\mathrm{e} c^2}{e} \simeq \frac{0.511~\mathrm{[MeV]}}{e}, \\
 d_0 & = & \frac{1}{k_{p}} \simeq \frac{5.32~\mathrm{\mu m}}{\sqrt{n_0}\mathrm{[10^{18}~cm^{-3}]}}, \\
 t_0 & = & \frac{1}{\omega_\mathrm{p}} \simeq \frac{17~\mathrm{fs}}{\sqrt{n_0 \mathrm{[10^{18}~cm^{-3}]}}}.
\end{eqnarray}

Our starting point is the master equation for the momenta of plasma electrons~\cite{bib:master1,bib:master2}:
\begin{equation}
\label{eq:master}
\frac{\partial^2 \mathbf{p}}{\partial t^2} + c^2 \nabla \times \nabla \times \mathbf{p} = -\left[\omega_{p}^2 + \frac{1}{m_\mathrm{e}}\nabla\cdot\left(\frac{\partial \mathbf{p}}{\partial t}+m_\mathrm{e} c^2\nabla \gamma \right)\right]\frac{\mathbf{p}}{\gamma} - m_\mathrm{e} c^2 \frac{\partial \nabla \gamma}{\partial t}.
\end{equation}
For a linearly polarized laser in $y$, and for plasma motion in $x$, the relevant components of Eq.~(\ref{eq:master}) are in $\emph{(x,y)}$ and read
\begin{eqnarray}
\frac{\partial^2 p_x}{\partial t^2} + \left(1+\frac{\partial^2}{\partial t \partial x}p_x + \frac{\partial^2 \gamma}{\partial x^2}\right)\frac{p_x}{\gamma} + \frac{\partial^2}{\partial t \partial x}\gamma & = & 0 \label{eq:masterx}, \\
\frac{\partial^2 p_y}{\partial t^2} - \frac{\partial^2 p_y}{\partial x^2} + \left(1+\frac{\partial^2}{\partial t \partial x}p_x + \frac{\partial^2 \gamma}{\partial x^2}\right)\frac{p_y}{\gamma} & = & 0 \label{eq:mastery}.
\end{eqnarray}

Since we are interested in the laser and plasma dynamics in the region that moves with the laser pulse we will also adopt the speed of light variables. This is a Galilean coordinate transformation to a frame that moves with the laser at $c$. We then transform time and space according to
\begin{eqnarray}
\label{eq:movingframe}
\psi & = & c t - x, \\
\tau & = & x/c,
\end{eqnarray}
where $\psi$ is a measure of the distance to the front of the laser and $\tau$ the time (or distance if multiplied by $c$) travelled by the laser pulse. The speed of light variables allow for the separation of the fast spatial scales associated with the variations in $\psi$ and that occur at the plasma wavelength $\lambda_\mathrm{p}$, from the slow laser temporal evolution associated with variations in $\tau$ and that scale with $\omega_0/\omega_\mathrm{p}\gg 1$. The latter scaling can be understood by noting that the Rayleigh length $k_\mathrm{p} Z_r$, which defines the typical time (distance) for the laser to evolve, is proportional to $\omega_0/\omega_\mathrm{p}$. In typical laser wakefield acceleration scenarios, where the plasma is transparent to the laser, $k_\mathrm{p} Z_r \propto \omega_0/\omega_\mathrm{p} \gg 1$. Hence, $\partial_{\tau}\simeq \omega_\mathrm{p}/\omega_0 \ll \partial_{\psi} \simeq 1/\lambda_\mathrm{p}$. We can therefore neglect $\partial_{\tau}$ in comparison to $\partial_{\psi}$. This is also referred to as the Quasi-Static Approximation (QSA)~\cite{bib:sprangle_prl_1990}. The QSA is valid as long as the laser pulse envelope does not evolve in the time it takes for an electron to go across the laser pulse. Under the QSA, the master equation components in $x$ and in $y$ given by Eqs.~(\ref{eq:masterx}) and (\ref{eq:mastery}) become
\begin{eqnarray}
\frac{\partial^2 p_x}{\partial \psi^2} + \frac{p_x}{\gamma}  \left(1-\frac{\partial^2 p_x}{\partial \psi^2} + \frac{\partial^2 \gamma}{\partial \psi^2}\right)- \frac{\partial^2 \gamma}{\partial \psi^2} & \simeq & 0 \label{eq:masterxnorm}, \\
2 \frac{\partial^2 p_y}{\partial \psi \partial \tau} + \frac{p_y}{\gamma}\left(1-\frac{\partial^2 p_x}{\partial \psi^2} + \frac{\partial^2 \gamma}{\partial \psi^2}\right) & \simeq & 0 \label{eq:masterynorm}.
\end{eqnarray}
Defining $\gamma - p_x = \chi$, we then arrive at the following simplified set of equations describing the coupled laser-plasma evolution and that can be used to describe wakefield excitation even in strongly non-linear regimes in one dimension:
\begin{eqnarray}
\left(\frac{p_x}{\gamma}-1\right)\frac{\partial^2 \chi}{\partial \psi^2} & = & -\frac{p_x}{\gamma}, \label{eq:_qsa_wake} \\
2\frac{\partial^2 p_y}{\partial \tau \partial \psi} + \left(1+\frac{\partial^2}{\partial \psi^2}\right)\frac{p_y}{\gamma}  & = & 0 \label{eq:_qsa_laser}.
\end{eqnarray}
Equation~(\ref{eq:_qsa_wake}) describes the wakefield evolution and Eq.~(\ref{eq:_qsa_laser}) the evolution of the laser (recall that $p_y = a_y$ due to the conservation of canonical momentum).

Equations~(\ref{eq:_qsa_wake}) and (\ref{eq:_qsa_laser}) have three unknowns, $p_y$, $p_x$ and $\chi$. In order to close the model we therefore need an additional relation between these quantities. This relation can be found by integrating Euler's equation. Since plasma waves are electrostatic in one dimension, Euler's equation becomes
\begin{equation}
\label{eq:euler}
\frac{\mathrm{d} p_x}{\mathrm{d} t} = - E_x - \frac{\partial \gamma}{\partial x} = \frac{\partial\left(\phi-\gamma\right)}{\partial x},
\end{equation}
where we have used $E_x = -\partial_{x} \phi$ in Eq.~(\ref{eq:euler}). Recasting Eq.~(\ref{eq:euler}) into the speed of light variables gives
\begin{equation}
\label{eq:euler_solv}
-\frac{\partial}{\partial \psi}\left(\gamma - p_x - \phi \right) = \frac{\partial}{\partial \tau}\left(\phi-\gamma \right) \simeq 0,
\end{equation}
where we have used $\partial_{\tau} \simeq 0$. As a result, for an electron initially at rest,
\begin{equation}
\label{eq:constant}
\chi = \gamma - p_x = 1 + \phi.
\end{equation}
Using the constant of motion given by Eq.~(\ref{eq:constant}), we can now relate $p_x$, $p_y$ and $\chi$ according to the following expressions:
\begin{eqnarray}
p_x & = & \frac{1+p_y^2-\chi^2}{2 \chi} \label{eq:px_1d} \\
\gamma & = & \frac{1+p_y^2+\chi^2}{2 \chi} \label{eq:gamma_1d}.
\end{eqnarray}
Equations~(\ref{eq:px_1d}) and (\ref{eq:gamma_1d}) can be used to close the non-linear model for the excitation of plasma waves given by Eqs.~(\ref{eq:_qsa_wake}) and (\ref{eq:_qsa_laser}) as in~\cite{bib:sprangle_prl_1990}:
\begin{eqnarray}
\frac{\partial^2 \chi}{\partial \psi^2} & = & -\frac{1}{2}\left(1-\frac{1+p_y^2}{\chi^2}\right), \label{eq:1d_qsa_wake} \\
2 \frac{\partial^2 p_y}{\partial \tau \partial \psi} + \frac{p_y}{\chi} & = & 0, \label{eq:1d_qsa_laser}
\end{eqnarray}
where $p_y = a_y$ is due to the conservation of the canonical momentum. Equations (\ref{eq:1d_qsa_wake}) and (\ref{eq:1d_qsa_laser}) are a system of non-linear coupled equations that describe wake excitation by ultra-intense and ultra-short laser pulses, valid in one dimension. They can also be referred to as 1D quasi-static equations. Equation~(\ref{eq:1d_qsa_wake}) describes wakefield excitation and Eq.~(\ref{eq:1d_qsa_laser}) describes the laser evolution. The source term for the laser evolution is $1/\chi$, and it is possible to show that $\chi$ is closely related to the plasma susceptibility. This can be shown with the help of the constant of motion given by Eq.~(\ref{eq:constant}) and by considering the continuity equation, which reads:
\begin{equation}
\label{eq:continuity}
\frac{\partial n }{\partial t} + \frac{\partial \left( n v_x\right)}{\partial z} = 0 \Rightarrow \frac{\partial\left[n\left(1-v_x\right)\right]}{\partial \psi} = \frac{\partial n}{\partial \tau}.
\end{equation}
Under the QSA Eq.~(\ref{eq:continuity}) can be integrated yielding $n(1-v_x) = n_0$. It is now possible to recover the 1D linear wakefield excitation model given in the previous sub-section by assuming $p_y\ll 1$ and $\chi \sim 1$.

The quasi-static equations (\ref{eq:1d_qsa_wake}) and (\ref{eq:1d_qsa_laser}) may be integrated analytically for specific laser pulse shapes assuming that the laser pulse remains unchanged. For a square shaped laser profile for instance, it is possible to show that the maximum wake potential $\phi_{\mathrm{max}}$ and maximum longitudinal momentum $p_{\mathrm{max}}$ of a plasma electron are respectively given by~\cite{bib:russian_qsa}
\begin{equation}
\label{eq:phimax}
\phi_{\mathrm{max}} \simeq \gamma_{\perp}^2-1,
\end{equation}
\begin{equation}
\label{eq:pmax}
p_{\mathrm{max}} \simeq \frac{\gamma_{\perp}^4-1}{2 \gamma_{\perp}^2},
\end{equation}
and that the maximum electric field of the plasma wave is
\begin{equation}
\label{eq:emax}
E_{\mathrm{max}} \simeq \frac{\gamma_{\perp}^2-1}{\gamma_{\perp}},
\end{equation}
where $\gamma_{\perp} = \sqrt{1+a_{y}^2}$. Equations~(\ref{eq:phimax}), (\ref{eq:pmax}) and (\ref{eq:emax}) show that higher intensity lasers lead to higher amplitude plasma waves with higher electron energy gain within the plasma wave. These equations cannot be used to explore particle trapping of background plasma electrons in the plasma wave (self-injection) because the QSA would no longer be valid in this case. Particle trapping occurs when the velocity of the plasma electrons matches the phase velocity of the plasma wave, which corresponds to the laser group velocity. For lasers with sufficiently high intensity, the velocity of the plasma electrons can be larger than the laser group velocity at the back of the first plasma wave. This leads to plasma wave-breaking. In the fluid approach that we have taken, wave-breaking leads to a singularity in the plasma density. For very high electron velocities, $\gamma \simeq p_x$, thus according to Eq.~(\ref{eq:constant}), $\phi \rightarrow -1$. Since $1/(1+\phi) = 1/\xi = n/\gamma$, then, as $\phi \rightarrow -1$, $n\rightarrow \infty$. In a kinetic description, wave-breaking corresponds to the crossing of electron trajectories. In relativistic regimes, the wave-breaking electric field limit is $e E_\mathrm{wb}/(m_\mathrm{e} c \omega_{p0}) = \sqrt{2}\sqrt{\gamma_{\phi}-1}$, where $v_{\phi}$ is the wake phase velocity and $\gamma_{\phi}=1/\sqrt{1-v_{\phi}^2}$ is the corresponding relativistic factor.

\begin{figure}[ht]
\begin{center}
\includegraphics[width=15 cm]{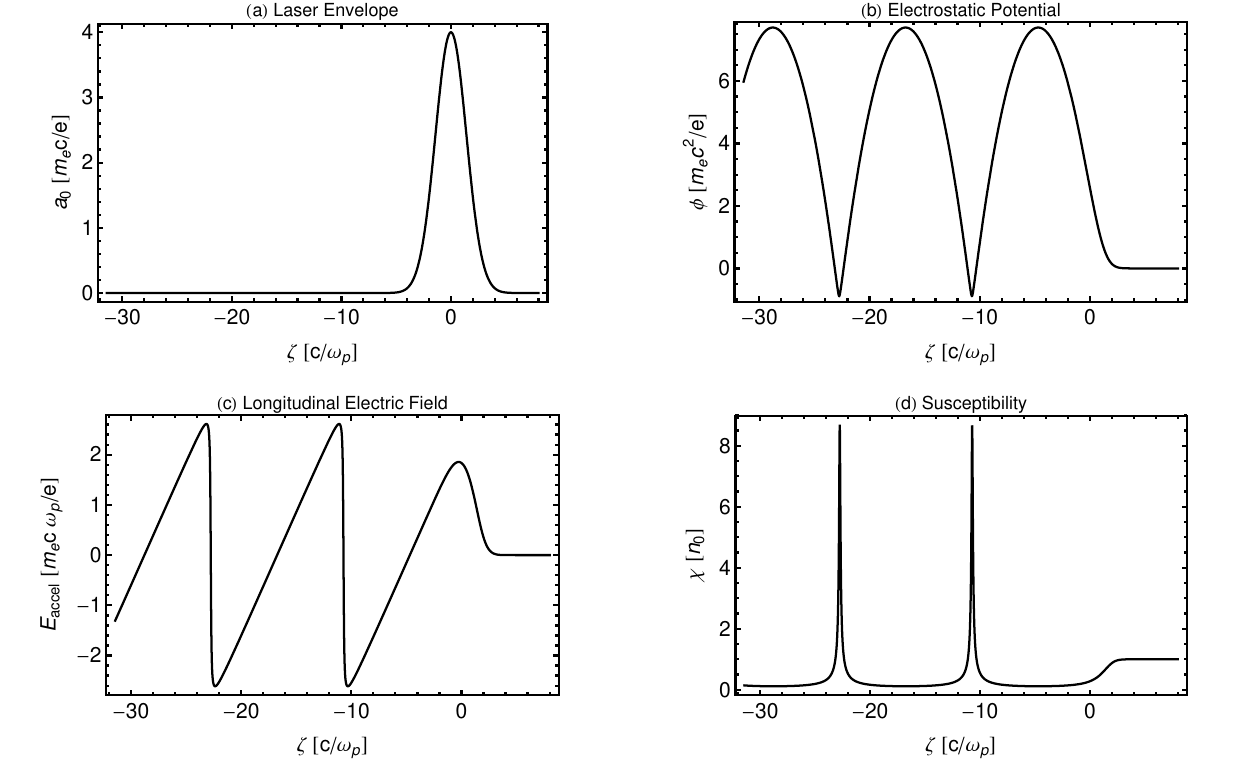}
\caption{Wakefield generation by an intense laser beam driver in one dimension: (a) shows the envelope of the laser vector potential; (b) shows the plasma electrostatic potential; (c) shows the longitudinal (accelerating) electric field; and (d) shows the plasma susceptibility.}
\label{fig:nonlinearwake1d}
\end{center}
\end{figure}

Key properties of 1D plasma wakefields excited by a short laser pulse in strongly non-linear regimes are shown in Fig.~\ref{fig:nonlinearwake1d}. Figure~\ref{fig:nonlinearwake1d} illustrates numerical solutions to the 1D QSA equations given by Eqs.~(\ref{eq:1d_qsa_wake}) and (\ref{eq:1d_qsa_laser}). Figure~\ref{fig:nonlinearwake1d}(a) shows the envelope profile of the laser pulse. The electrostatic potential, shown in Fig.~\ref{fig:nonlinearwake1d}(b) is strongly non-linear, i.e. $\phi$ is not well described by sinusoidal oscillations at $\lambda_\mathrm{p}$ containing additional high-order harmonics. In addition, the minimum $\phi$ approaches $\phi \rightarrow -1$, as predicted theoretically using the constant of motion. Another distinctive feature of non-linear wake excitation is illustrated in Fig.~\ref{fig:nonlinearwake1d}, which shows a typical sawtooth shape for the accelerating electric field. The plasma susceptibility, and thus the plasma density, also reach very high values at the end of the first plasma wave, where the longitudinal electron momentum is at a maximum, which is also consistent with the theoretical arguments provided in the previous paragraph.

It is important to note that despite not being able to capture particle trapping in plasma waves, the QSA can be used to explore wakefield excitation and the acceleration of externally injected particle bunches in the plasma wave fields. The computational advantages of the QSA when compared to kinetic descriptions (such as the PIC method), have led to the development of reduced numerical models. These reduced models are used to predict the output of plasma-based experiments. Examples of such codes are QuickPIC~\cite{bib:quickpic}, WAKE~\cite{bib:wake} and HiPACE~\cite{bib:desy_qsa}.

\subsection{Beam loading in the linear regime}

In order to explore the acceleration of particle bunches to maximize the charge that can be accelerated and to minimize the final energy spreads, it is important to investigate how external charged particle bunches influence the wakefield structure. This research topic is typically called beam loading. The study of beam loading in plasma waves is thus important to shape the current profile of the particle bunches for maximizing the quality of the accelerated particles which is critical to potential applications. In this section, we will outline important results on the beam loading of electron bunches in one dimension and in the linear regime.

\begin{figure}[ht]
\begin{center}
\includegraphics[width=8.1 cm]{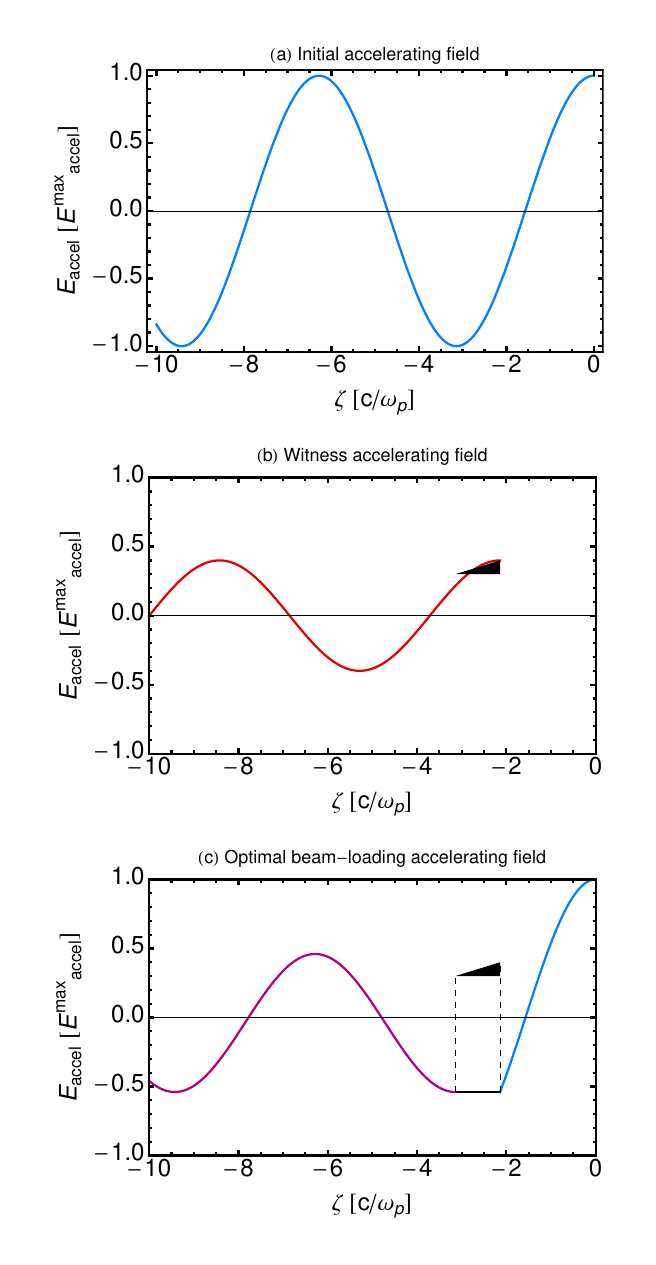}
\caption{Illustration of optimal beam loading conditions to minimize the energy spread of an energetic electron beam (dark triangle): (a) shows the accelerating gradient of a plasma wave driven by a laser pulse in the linear regime; (b) shows the accelerating wakefield driven by a witness electron bunch; and (c) shows the accelerating wakefield resulting from the combination (sum) of wakefield in (a) and in (b).}
\label{fig:beamloading1d}
\end{center}
\end{figure}

In Fig.~\ref{fig:beamloading1d} the optimal beam loading conditions in the linear regime and in one dimension are illustrated. Figure~\ref{fig:beamloading1d}(a) shows a portion of the initial accelerating field driven by a laser pulse driver in the linear regime. Figure~\ref{fig:beamloading1d}(b) shows the wakefield contribution driven by a witness electron bunch with a triangular shape. The resulting wakefield (Fig.~\ref{fig:beamloading1d}(c)) shows that the acceleration gradient is constant throughout the entire witness bunch. The witness bunch then decreases the absolute value of the electric field in the beam region, thereby reducing the acceleration gradient. As a result, all beam electrons will accelerate with similar accelerating fields. This configuration then preserves the initial witness bunch energy spread, yielding a scenario of ideal beam loading.

Although a more refined beam loading theory can be derived in the linear regime for witness bunches with arbitrary shapes, we will adopt a simple model assuming a very short, uniform density witness bunch in comparison to the plasma wavelength.

We assume that the bunch density is $n_\mathrm{b}$. An estimate for the maximum number of particles ($N_0$) that can be loaded into the wakefield in one dimension and in the linear regime can be found by assuming that the total electric field vanishes at the location of the beam. This condition leads to~\cite{bib:katsouleas_bemloading}
\begin{equation}
\label{eq:maxcharge_1d}
N_0 = 5\times 10^5 \left(\frac{n_\mathrm{b}}{n_0}\right) \sqrt{n_0} A,
\end{equation}
where A is the area of the transverse bunch section and $n_\mathrm{b}$ is the trailing bunch density. The efficiency is $100\%$ if the total number of particles loaded into the wakefield matches Eq.~(\ref{eq:maxcharge_1d}). However, this comes at the expense of obtaining $100\%$ energy spread because the front of the beam accelerates at the maximum gradient, while the back of the beam does not accelerate. Since the energy gain $\Delta \gamma \propto E_{\mathrm{accel}}$ (where $E_{\mathrm{accel}}$ is the accelerating gradient) the final energy spread when the number of particles ($N$) in the bunch is lower than $N_{0}$ at:
\begin{equation}
\label{eq:energyspread}
\frac{\Delta \gamma_{\mathrm{max}}-\Delta \gamma_{\mathrm{min}}}{\Delta \gamma_{\mathrm{max}}} = \frac{E_\mathrm{i}-E_\mathrm{f}}{E_\mathrm{i}} = \frac{N}{N_0},
\end{equation}
where $\Delta \gamma_{\mathrm{max/min}}$ are the maximum/minimum energy gains by the bunch particles and $E_\mathrm{i/f}$ are the accelerating electric fields at the front/behind of the accelerating bunch. Equation~(\ref{eq:energyspread}) illustrates the trade-off between number of accelerated particles and final energy spread. Hence, maximizing the accelerated charge also maximizes the energy spread and \emph{vice versa}.

Another key feature related with beam loading is the fraction of the energy absorbed by the trailing particle bunch in the wakefield. Since the energy of the wakefield scales with $\propto E_{\mathrm{accel}}^2$, the fraction of energy absorbed by the witness particle bunch is $1-E_\mathrm{f}^2/E_\mathrm{i}^2$. Using Eq.~(\ref{eq:energyspread}), $E_\mathrm{f} = E_\mathrm{i}(1-N/N_0)$. Thus, the energy conversion efficiency from the wakefield to a trailing bunch of particles is
\begin{equation}
\label{eq:efficiency-kg}
\eta_\mathrm{b} = \frac{N}{N_0}\left(2-\frac{N}{N_0}\right).
\end{equation}
Equations~(\ref{eq:energyspread}) and (\ref{eq:efficiency-kg}) also illustrate the trade-off between the beam loading efficiency and the final bunch energy spread. Small energy spreads require smaller total beam charges which leads to lower energy conversion efficiencies. Higher energy conversion efficiencies however lead to higher energy spreads. Although these conclusions are generally valid, it is possible to derive conditions for optimal beam loading using shaped current density profiles, as Fig.~\ref{fig:beamloading1d} shows.

\section{Strongly non-linear plasma waves in multidimensions: the blowout regime}
\label{sec:fenoblowout}

The wakefield excitation models derived in the previous Section, valid in one dimension, are useful in describing laser wakefield accelerators qualitatively. However, because in typical experiments the driver has transverse dimensions smaller or comparable to the plasma wavelength, the 1D theory cannot be employed to describe and predict the experimental outputs quantitatively. In typical experiments, the structure of the wakefields is inherently multidimensional. An accurate understanding of the experiments therefore requires the inclusion of multidimensional effects in both the theory and simulations.

Multidimensional plasma oscillations are intrinsically non-linear and anharmonic even for low-amplitude plasma waves~\cite{bib:dawson_pr_1959}. The period of plasma electron oscillations then depends on their amplitude of oscillation. Although for low-amplitude plasma waves anharmonic effects can be neglected for the first few plasma waves, the cumulative phase shift between close electrons becomes substantial when the number of plasma oscillations increases. As a result, the trajectory of adjacent electrons will always cross in multidimensional plasma waves when close electrons become $\pi/2$ out of phase. Trajectory crossing inevitably leads to wave-breaking, where the initial laminar electron flow becomes turbulent. This is in stark contrast with the 1D theory developed in the previous sections where the flow remains laminar as long as it is laminar during the first plasma oscillations. When trajectory crossing occurs and the electron flow becomes turbulent, the plasma fluid equations used in the previous Section cease to be valid. The properties of the wakefield in these scenarios can only be captured by kinetic descriptions.

When the wakefield amplitude is large, sheath crossing can occur during the first plasma wave. Current experiments operate in this strongly non-linear regime. In this section, we will outline the derivation of a non-linear model for the wakefield excitation in strongly non-linear regimes and in multidimensions which is capable of describing and predicting experimental outputs.

Although we will focus the analysis on laser wakefield acceleration, general conclusions from this section are also valid for non-linear wakefields driven by electron beam drivers because the main properties of the wakefield in strongly non-linear regimes are nearly independent of the nature of the driver. Positron beams can also be used to drive strongly non-linear plasma waves and many of the findings outlined in this section may also be applicable to this case. However, we note that the wakefield generation process can differ significantly from that of lasers or electron bunches. While electron beams or laser pulses repel plasma electrons from the region of the driver, positrons attract plasma electrons towards the axis~\cite{bib:lee_pre_2001}.

\begin{figure}[ht]
\begin{center}
\includegraphics[width=15 cm]{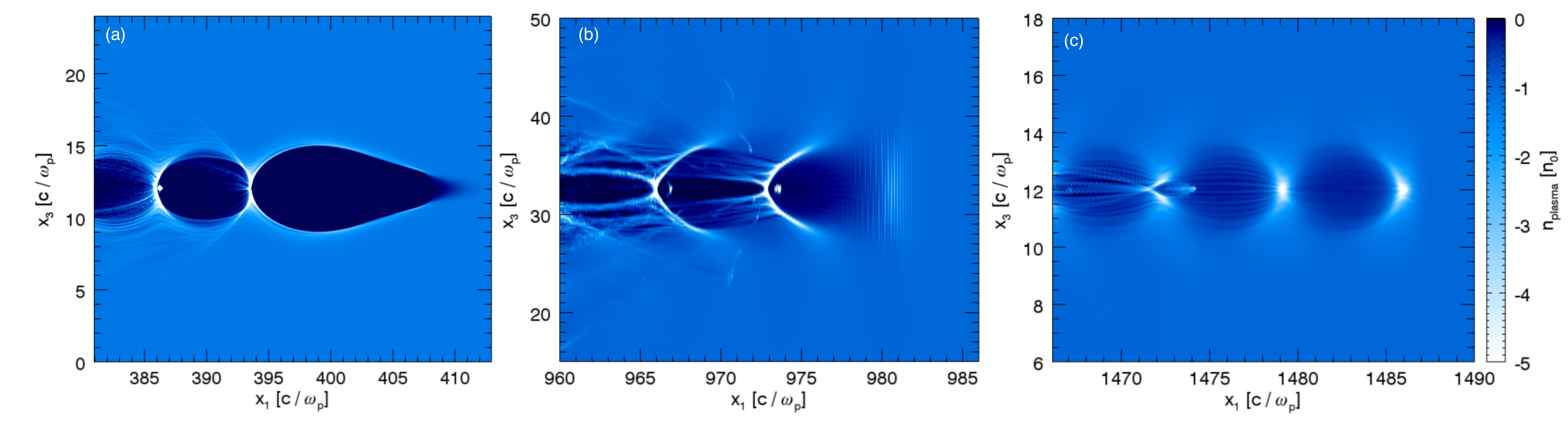}
\caption{OSIRIS simulation results illustrating the generation of strongly non-linear plasma waves. The colours are proportional to the electron plasma density. The driver moves from left to right: (a) shows the wakefield driven by an ultra-relativistic particle beam driver; (b) shows the generation of wakefields driven by intense laser beams; and (c) shows the plasma density perturbation associated with a positron bunch driver.}
\label{fig:3dblowout}
\end{center}
\end{figure}

Figure~\ref{fig:3dblowout} shows a particle-in-cell OSIRIS~\cite{bib:osiris,bib:fonseca_ppcf_2013} simulation results of a strongly non-linear wakefield excitation in multidimensions driven by an intense electron bunch (Fig.~\ref{fig:3dblowout}(a)), laser pulse (Fig.~\ref{fig:3dblowout}(b)) and positron bunch (Fig.~\ref{fig:3dblowout}(c)). A distinctive signature corresponding to the excitation of strongly non-linear plasma waves in multidimensions is electron cavitation, {i.e.} the generation of a region void of plasma electrons. This regime is called the blowout~\cite{bib:lu_prl_2006} or bubble regime~\cite{bib:pukhov_apl_2002} when electrons (Fig.~\ref{fig:3dblowout}(a)) or laser pulse (Fig.~\ref{fig:3dblowout}(b)) drivers are employed, and suck-in regime when positron bunch drivers (Fig.~\ref{fig:3dblowout}(c)) are used ~\cite{bib:lee_pre_2001}. Although the wakefields share many similarities, and the main properties of the wakefield are driver independent, Fig.~\ref{fig:3dblowout} shows that finer details of the blowout region depends on the nature of the driver. Figures~\ref{fig:3dblowout} also show electron bunches [in the second bucket in Fig.~\ref{fig:3dblowout}(a) and Fig.~\ref{fig:3dblowout}(c) and in the first and second buckets in Fig.~\ref{fig:3dblowout}(b)] in conditions to be
accelerated by the plasma. 

For the remainder of this report, we define the transverse size of the laser as the laser spot-size $W_0$. The laser duration is $\tau_\mathrm{L}$. The peak vector potential $a_0$ is related to the laser intensity through $a_0 \simeq 0.8 \left(\lambda_0/1 [\mathrm{\mu m}]\right) \left[I/\left(10^{18}\mathrm{W\,cm^{-2}}\right)\right]^{1/2}$, where $\lambda_0=2 \pi c /\omega_0$ is the central laser wavelength and $\omega_0$ is the central laser frequency.

When an intense laser or electron bunch driver propagates through a plasma it radially expels plasma electrons away at its passage (see Fig.~\ref{fig:3dblowout}). If the laser intensity (electron driver density) is sufficiently high, the driver expels nearly all plasma electrons away from the region in which the driver propagates. Electron trajectories cross and accumulate in a thin, high-density electron layer that surrounds an electron void. Ions remain stationary and push plasma electrons back to the axis after the driver has propagated a distance close to $\sim \lambda_\mathrm{p}$. At the back of the plasma wave, the shape of the thin electron layer resembles a sphere or a bubble.

At the back of the sphere, there are strong electron accelerating fields which can be sufficiently strong to capture a fraction of the background plasma electrons into the bubble (self-injection). These electrons can then be focused and accelerate to high energies in the focusing and accelerating fields of the bubble. In the laser wakefield accelerator, reaching these strongly non-linear regimes requires normalized laser vector potentials close to $a_0\gtrsim 2$ for spot-sizes ($W_0$) of a few skin-depths. In the plasma (beam driven) wakefield accelerator, the blowout regime requires that particle bunch densities are larger than the background plasma density.

\subsection{Phenomenological model for the blowout regime}
In order to describe electron acceleration in non-linear laser driven wakefields we first focus on a phenomenological model for the blowout regime driven by laser pulses~\cite{bib:lu_prstab_2007}. Our first goal is first to derive a set of scaling laws for stable laser wakefield acceleration and which are capable of predicting key output beam properties such as energy and charge.

Important wakefield properties are defined by the focusing and accelerating fields. Focusing fields are absent from 1D descriptions but are crucial to the prediction of the outputs of the experiments. The focusing field for a relativistic charged particle traveling at $c$ accelerating in the blowout region is then given by
\begin{equation}
\label{eq:focusing}
W_{\perp} = E_r - B_{\theta} = \frac{r}{2},
\end{equation}
where $r$ is the distance to the axis, $E_r$ is the radial electric field and $B_{\theta}$ the azimuthal magnetic field. Equation~(\ref{eq:focusing}) shows that the field is always focusing for electrons. Linear focusing fields are a feature of strongly non-linear regimes which are absent from wakefield excitation in the linear regime. As the electrons accelerate, they perform transverse harmonic oscillations (also called betatron oscillations). This is important to preserve beam emittance as the acceleration progresses. Achieving small (as small as possible) beam emittances is crucial for high-energy physics and radiation applications. Hence linear focusing fields are crucial for potential applications of plasma accelerators. In the linear regime, the focusing fields are non-linear with high-order contributions in powers of $r$. Thus, unless the electron bunches are much narrower than the driver spot-size so that they are subject to linear focusing fields near the axis, they will perform anharmonic oscillations, which will lead to higher final bunch emittances.

\begin{figure}[ht]
\begin{center}
\includegraphics[width=12 cm]{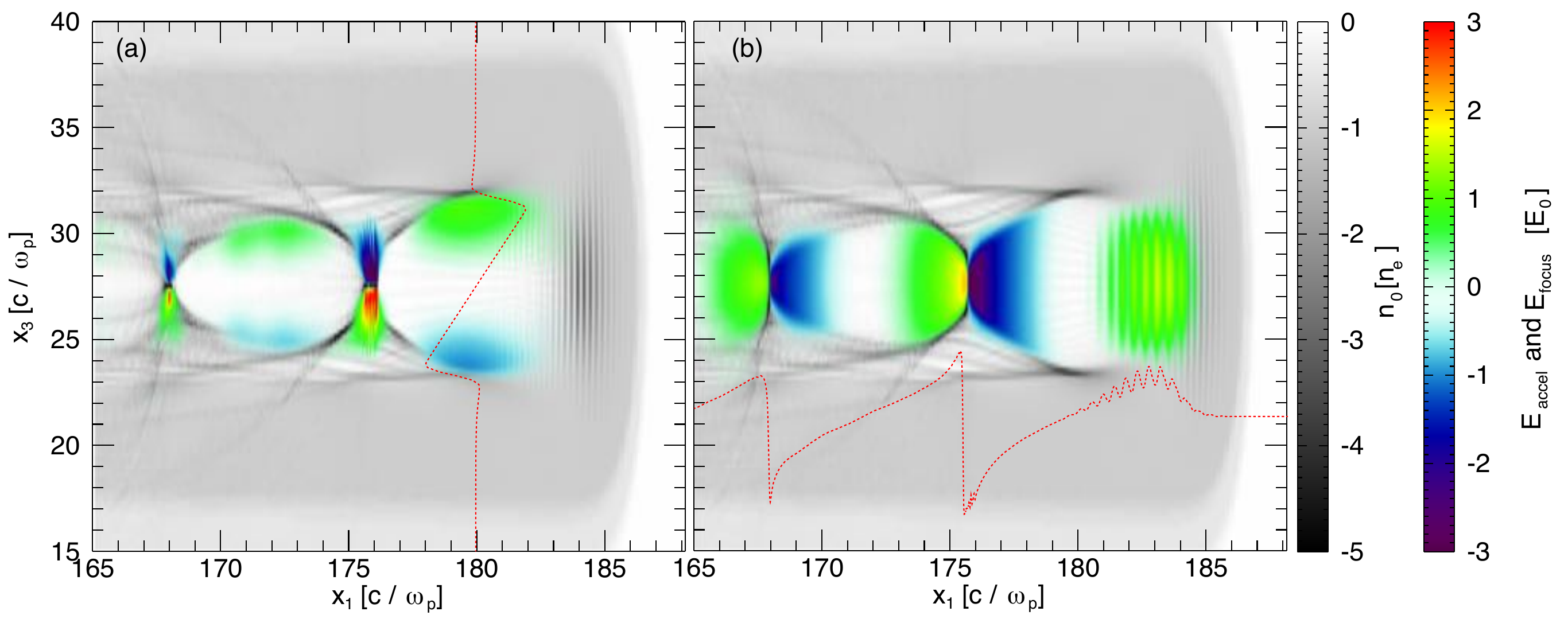}
\caption{Three-dimensional (3D) Osiris simulation showing focusing (a) and accelerating fields (b) for a laser wakefield accelerator in the blowout regime. The plasma density is shown in grey colours and fields in blue-red colours. The red dashed lines shown a transverse lineout of the focusing (a) and accelerating (b) fields. Focusing fields are linear within the entire blowout region. Accelerating fields are close to linear except at the back of the bubble.}
\label{fig:lwfablowout}
\end{center}
\end{figure}

The accelerating field determines the maximum energy of a witness electron bunch. The accelerating electric field felt by the bunch electrons in the bubble is given by
\begin{equation}
\label{eq:accelerating}
E_{\mathrm{accel}} = \frac{\xi}{2},
\end{equation}
where $\xi$ measures the distance to the centre of the bubble which moves nearly at the laser group velocity. The field accelerates electrons for $\xi<0$ and decelerates them for $\xi>0$. The linear focusing and accelerating fields are well reproduced in simulations of the laser wakefield accelerator, as shown in Fig.~\ref{fig:lwfablowout}.

Equation~(\ref{eq:accelerating}) is valid as long as the witness beam wakefields are negligible. In this case, Eq.~(\ref{eq:accelerating}) shows that the acceleration depends on the initial $\xi$ position of each beam electron. This can increase the final energy spread of the bunch. As with the 1D case, however, it is possible to compensate for this effect by tailoring the shape of the witness bunch current profile in order to preserve the initial energy spread of the witness bunch.

In order to find the average accelerating field in the bubble, which when multiplied by the total acceleration distance determines the final energy energy gain, Eq.~(\ref{eq:accelerating}) needs to be supplemented with an additional expression defining the radius of the bubble. The radius of the bubble, $r_\mathrm{b}$, can be retrieved by equating the laser ponderomotive (repulsive) force to the ion channel (attractive) force:
\begin{equation}
\label{eq:rb}
F_\mathrm{p} \simeq E_r \Leftrightarrow \frac{a_0}{w_0} = E_r \simeq r_\mathrm{b} \rightarrow r_\mathrm{b} = \alpha \sqrt{a_0},
\end{equation}
where $\alpha=2$ has been determined through PIC simulations~\cite{bib:lu_prl_2006,bib:lu_prstab_2007} and where we have assumed that $w_0 \simeq r_\mathrm{b}$. Combining Eq.~(\ref{eq:accelerating}) with Eq.~($\ref{eq:rb}$) yields an estimate for the average accelerating field given by
\begin{equation}
\label{eq:accelerating_ave}
\langle E_{\mathrm{accel}} \rangle \simeq \frac{\sqrt{a_0}}{2}.
\end{equation}

The maximum acceleration distance corresponds to the smallest distance between pump depletion or dephasing. The pump depletion length, $L_{\mathrm{pd}}$, is the length it takes for the laser to exhaust its energy to the plasma through wakefield excitation. For propagation distances larger than $L_{\mathrm{pd}}$, the amplitudes of the plasma waves are negligible. Thus, we can assume that the acceleration stops at $L_{\mathrm{pd}}$. The dephasing length, $L_{\mathrm{d}}$, is the length it takes for a particle to outrun the accelerating phase of the wave, {i.e.} to go from regions with $\xi<0$, where $E_{\mathrm{accel}}<0$, to regions with $\xi=0$ where $E_{\mathrm{accel}}=0$.

Pump depletion in the blowout regime is determined by the rate at which the laser leading edge gives its energy to the plasma. This localized pump depletion process is also called etching. Since the back propagates mostly in vacuum, it does not give energy to the plasma. As the laser propagates, the front of the laser is then locally pump depleted. The pump depletion length is then given by the product between the laser duration and the velocity at which the laser leading edge etches back, given by $v_{\mathrm{etch}}=c \omega_\mathrm{p}^2/\omega_0^2$~\cite{bib:decker_pop_1996}
\begin{equation}
\label{eq:Lpd}
L_{\mathrm{pd}} = \frac{\omega_0^2}{\omega_\mathrm{p}^2} \left(c \tau_L\right).
\end{equation}

The maximum dephasing length is given by the length it takes for a particle travelling at $c$ to outrun the accelerating phase of the wakefield traveling with a phase velocity $v_{\phi}$. For an electron initially at $\xi=r_\mathrm{b}$, $ L_d = c r_\mathrm{b}/(c-v_{\phi})$. Since the wake phase velocity is $v_{\phi}= v_g-v_{\mathrm{etch}}$, where $v_g$ is the laser linear group velocity given by
\begin{equation}
\label{eq:vg_linear}
v_g = \frac{\partial \omega}{\partial k} = 1 - \frac{1}{2}\frac{\omega_0^2}{\omega_\mathrm{p}^2},
\end{equation}
the dephasing length is
\begin{equation}
\label{eq:Ld}
\frac{(c-v_{\phi})}{c} L_\mathrm{d} = r_\mathrm{b} \Leftrightarrow L_\mathrm{d} = \frac{2}{3}\frac{\omega_0^2}{\omega_\mathrm{p}^2} r_\mathrm{b}.
\end{equation}

Combining Eqs.~(\ref{eq:Lpd}) and (\ref{eq:vg_linear}) yields a criteria for choosing the laser duration for optimal acceleration such that no laser energy is left after the electrons outrun the wave at $L_{\mathrm{pd}}=L_{\mathrm{d}}$:
\begin{equation}
\label{eq:tau}
\tau_L = \frac{2}{3} r_\mathrm{b}.
\end{equation}

We can now estimate the maximum energy $\Delta E = q \langle E_{\mathrm{accel}}\rangle L_{\mathrm{accel}}$ gained by an electron in the blowout regime. Denoting the acceleration distance by $L_{\mathrm{accel}}=L_{\mathrm{d}}=L_{\mathrm{pd}}$ then
\begin{equation}
\label{eq:energygain}
\Delta E = \frac{2}{3} m_\mathrm{e} c^2 \frac{\omega_0^2}{\omega_\mathrm{p}^2} a_0.
\end{equation}

So far, we have neglected the influence of the transverse laser dynamics on wakefield excitation and electron acceleration. This approximation is valid as long as the laser propagation and wakefields remain stable during $L_{\mathrm{accel}}$. In order to stabilize the transverse laser dynamics, we need to explore how to prevent laser Rayleigh diffraction, one of the key processes that can degrade wakefield excitation and electron acceleration. Theory, simulations and experiments have shown that plasmas can act as optical fibers, guiding the propagation of intense lasers over distances that largely exceed the Rayleigh length. In strongly non-linear regimes, the blowout region refractive index gradients are sufficient to self-guide the body of the driver. Through simulations, it has been found that the optimal condition for stable, self-guided laser propagation occurs when $W_0 = r_\mathrm{b} = 2 \sqrt{a_0}$ as long as $a_0>2$~\cite{bib:lu_prl_2006,bib:lu_prstab_2007}. The laser front, which propagates in a region of nearly undisturbed plasma, may still diffract. This can be avoided if the etching rate exceeds the diffraction rate. This condition is met when
\begin{equation}
\label{eq:selfguiding}
a_0 \gtrsim \left(\frac{n_\mathrm{c}}{n_\mathrm{p}}\right)^{1/5}.
\end{equation}
Equation~(\ref{eq:selfguiding}) is generally valid for $a_0\gtrsim 4$. For $a_0\gtrsim 2$, an external parabolic plasma channel needs to be present to externally guide the laser pulse.

In addition to determining maximum accelerating gradients and final energies in the blowout regime, the accelerating and focusing wakefields given by Eqs.~(\ref{eq:focusing}) and (\ref{eq:accelerating}) also define important beam loading properties such as the maximum charge that can be accelerated. To estimate the maximum amount of accelerated charge, we assume that a witness electron bunch absorbs all the energy contained in the longitudinal and focusing bubble fields. The electromagnetic energy of the wakefield in the blowout regime is
\begin{equation}
\label{eq:fieldene}
\varepsilon_{\|} \simeq \varepsilon_{\perp} \simeq \frac{1}{120} \left(k_\mathrm{p} r_\mathrm{b}\right)^5 \left(\frac{m_\mathrm{e}^2 c^5}{\omega_\mathrm{p} e^2}\right),
\end{equation}
and the energy absorbed by $N$ particles, assuming an average accelerating field gradient $E_{\mathrm{accel}}\simeq r_\mathrm{b}/2$, is
\begin{equation}
\label{eq:partene}
\varepsilon_{e^{-}} \simeq m_\mathrm{e} c^2 N\left(\frac{k_\mathrm{p} r_\mathrm{b}}{2}\right)^2.
\end{equation}
Matching Eq.~(\ref{eq:fieldene}) to Eq.~(\ref{eq:partene}) then gives
\begin{equation}
\label{eq:maxpart}
N \simeq \frac{1}{30} \left(k_\mathrm{p} r_\mathrm{b}\right)^3 \frac{1}{k_\mathrm{p} r_\mathrm{e}},
\end{equation}
where $N$ is the maximum number of electrons that can be loaded into the wakefield. The acceleration efficiency is the fraction of laser energy that goes into the accelerated electrons. Since the laser energy scales with $r_\mathrm{b}^3 a_0^2$ (assuming $W_0 \simeq c\tau_L  \simeq r_\mathrm{b}$), then the efficiency goes as
\begin{equation}
\label{eq:efficiency}
\Gamma \simeq \frac{1}{a_0}.
\end{equation}

Equations~(\ref{eq:energygain}), (\ref{eq:maxpart}) and (\ref{eq:efficiency}) illustrate the trade-off between energy gain, maximum number of accelerated particles and efficiency. For a constant laser energy, lower laser $a_0$s leads to higher efficiencies at the expense of lower accelerated charge and longer accelerating distances that result in final higher energies. Higher laser $a_0$s lead to lower efficiencies, lower final bunch energies, but to higher charge. In addition, the acceleration distance is also smaller for higher $a_0$.

The scaling laws derived above can also be rewritten in practical units as
\begin{eqnarray}
\tau \mathrm{[fs]}& = & 53.22\left(\frac{\lambda_0\left[\mu \mathrm{m}\right]}{0.8}\right)^{2/3}\left(\frac{\epsilon\mathrm{[J]}}{a_0^2}\right)^{1/3} \label{eq:matchedlength}, \\
w_0 & = & \frac{3}{2}c \tau_L \label{eq:matchedspot}, \\
n_0\left[\mathrm{cm}^{-3}\right] & \simeq & 3.71 \frac{a_0^3}{P\mathrm{\left[TW\right]}} \left(\frac{\lambda_0 [\mathrm{\mu m}]}{0.8}\right)^{-2} \label{eq:matcheddensity}.
\end{eqnarray}
The total acceleration distance, final energy and maximum accelerated charge are:
\begin{eqnarray}
L_{\mathrm{accel}} \mathrm{[cm]} & \simeq & 14.09\frac{\epsilon \mathrm{[J]}}{a_0^3} \label{eq:matcheddistance}, \\
\Delta E \mathrm{[GeV]} & \simeq & 3 \left(\frac{\epsilon \mathrm{[J]}}{a_0^2}\frac{0.8}{\lambda_0 \mathrm{[\mu m]}}\right)^{2/3} \label{eq:matchedenergy}, \\
q\mathrm{[nC]} & \simeq & 0.17 \left(\frac{\lambda_0 [\mathrm{\mu m}]}{0.8}\right)^{2/3} \left(\epsilon\mathrm{[J]} a_0\right)^{1/3} \label{eq:matchedcharge}.
\end{eqnarray}

\begin{figure}[ht]
\begin{center}
\includegraphics[width=13.8cm]{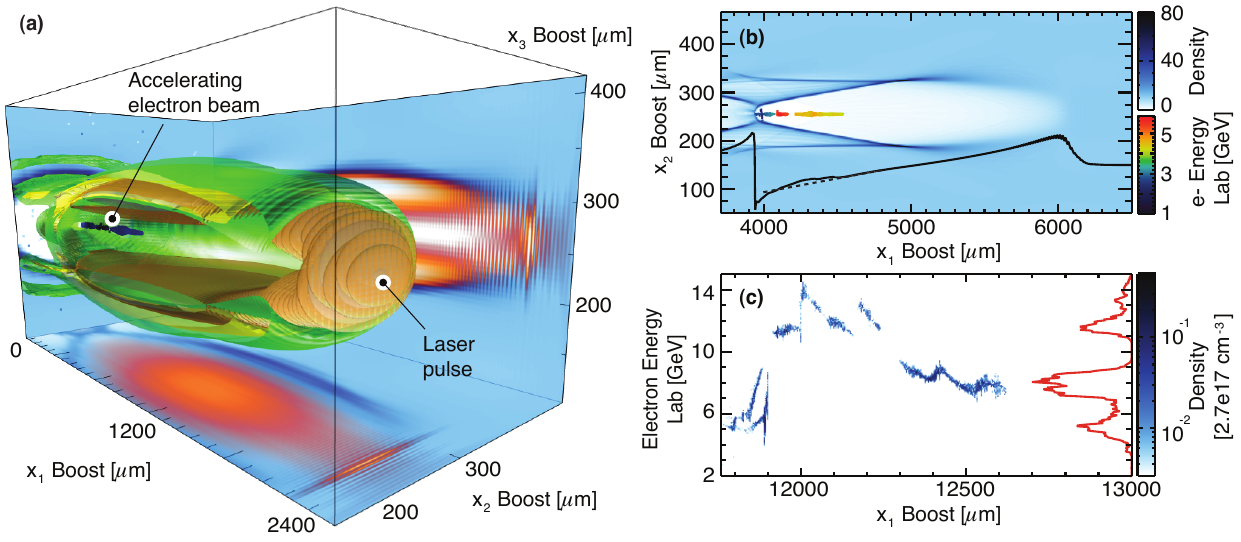}
\caption{(Picture taken from Ref.~\cite{bib:martins_np_2010}) 3D PIC simulation in a Lorentz boosted frame of a laser wakefield accelerator using a 250 J laser pulse in self-injection, self-guiding scenarios: (a) shows the laser pulse in orange, the accelerated beam in dark blue and plasma density isosurfaces in green and yellow. Blue projections represent the background electron plasma density and orange projections the focusing and laser pulse electric fields; (b) shows a central slice of the simulation box illustrating the plasma electron density (blue) and self-injected electrons coloured according to their energy. The line-out (black) represents the longitudinal electric field ($E_{\mathrm{accel}}$). The dashed line (black) represents the theoretical prediction (Eq.~(\ref{eq:accelerating})); and (c) represents the phase-space of self-injected electrons. The red line represents the integrated energy spectrum.}
\label{fig:elisi}
\end{center}
\end{figure}

These scalings have been used to guide and predict the output of current laser wakefield acceleration experiments, and to guide the design of future experiments using some of the most powerful lasers soon to become available. They have also been confirmed through numerous 3D PIC simulations performed with different algorithms. For example, PIC simulations performed in relativistic boosted frames have illustrated the acceleration of 12--14 GeV electron bunches with 1--2 nC each (${\sim}4.8$ nC in total) using lasers that will soon become available at the Extreme Light Infrastructure (ELI)~\cite{bib:martins_np_2010}. Simulations showed that the acceleration could take place over distances smaller than 10 cm. These simulations were performed for strongly non-linear regimes using $a_0=4$, where a fraction of the background plasma electrons were trapped and accelerated to these high energies. These results are illustrated in Fig.~\ref{fig:elisi}.

\begin{figure}[ht]
\begin{center}
\includegraphics[width=15 cm]{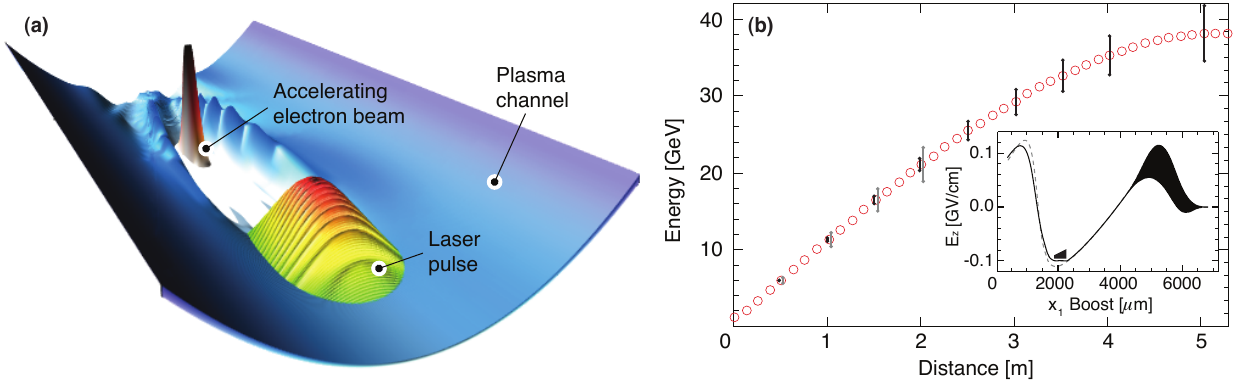}
\caption{Picture taken from Ref.~\cite{bib:martins_np_2010}) 3D PIC simulation results in a Lorentz boosted frame of a laser wakefield accelerator using a 250 J laser pulse in external injection and external guiding scenarios: (a) illustration of the parabolid plasma channel used to guide the laser pulse (blue). The laser pulse is also shown. An externally injected electron bunch accelerates at the back of the plasma wave; and (b) shows the evolution of the maximum energy as a function of the acceleration distance. The vertical line segments correspond to the energy spread of the bunch. The inset shows the relative position of the externally injected bunch and longitudinal accelerating field. The bunch current profile was designed in order to lead to a flat accelerating field in the region of the bunch to minimize the growth of the energy spread.}
\label{fig:eliei}
\end{center}
\end{figure}

Simulations have also been performed at lower laser intensities with $a_0=2$, where self-injection is absent, and using laser energies close to 250 J. The acceleration of an externally injected electron bunch was then investigated. Simulations showed the acceleration of a 40 GeV electron beam with 0.3~nC in a preformed parabolic plasma channel 5 m long. These results are shown in Fig.~\ref{fig:eliei}.

\begin{figure}[ht]
\begin{center}
\includegraphics[width=15 cm]{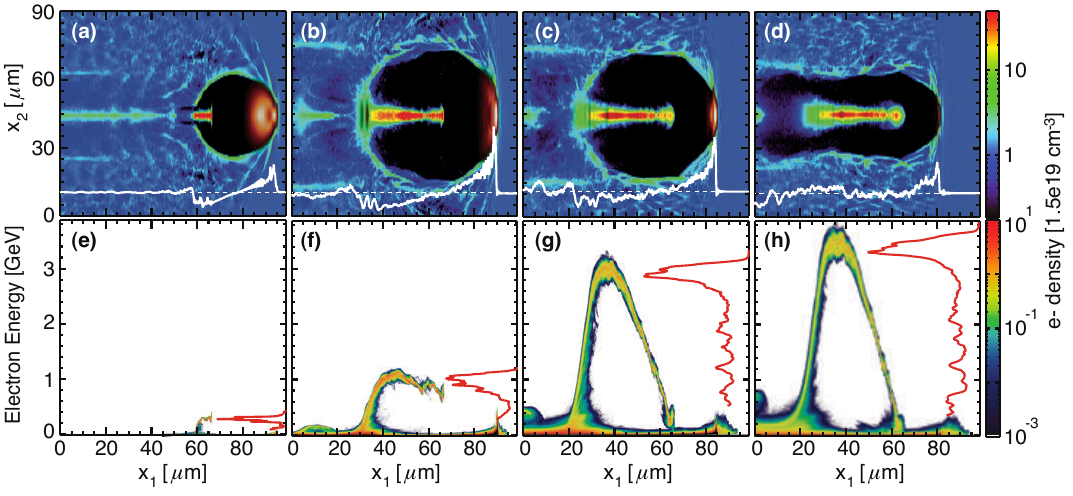}
\caption{Picture taken from Ref.~\cite{bib:martins_np_2010}). 3D simulation results illustrating a laser wakefield accelerator in the bubble regime using a laser with $a_0=53$: (a)--(d) show central slices of the simulation box illustrating the plasma density (colours). On-axis accelerating field lineouts are represented in white; and (e)--(h) show the phase space of the plasma (colours) and integrated energy spectrum (red line).}
\label{fig:gordienko}
\end{center}
\end{figure}

The scalings presented here are strictly valid for $2\lesssim a_0 \lesssim 2 \left(\omega_0/\omega_\mathrm{p}\right)^{1/4}$. However, electron acceleration can also occur at much higher laser intensities. For instance, using $a_0=53$, 3 GeV electron bunches with high charges of around 25 nC could be achieved~\cite{bib:gordienko_pop_2004}, although at the expense of higher energy spreads. These results are illustrated in Fig.~\ref{fig:gordienko}.

\section{Theory for the blowout regime}
\label{sec:blowout}

In this section, we will outline an analytical derivation for the electromagnetic field structure of the wakefield in the blowout regime~\cite{bib:lu_prl_2006}. The theory assumes cylindrical symmetry and employs the quasi-static approximation. Thus, as with to the 1D scenario (cf. Section~\ref{sec:nonlinear1d}), the motion of plasma electrons in multidimensions is also characterized by a constant of motion relating the particle velocity to the wake potential (cf. Eq.~(\ref{eq:constant})). This constant of motion plays a very important role in understanding the trapping process in multidimensions and also in defining the shape of the blowout region. In order to generalize Eq.~(\ref{eq:constant}) for the multidimensional scenario, we then start by considering the Hamiltonian of a charged particle in an electromagnetic field:
\begin{equation}
\label{eq:hamiltonean}
H = \sqrt{m_\mathrm{e}^2 c^4+\left(\mathbf{P}+\frac{e \mathbf{A}}{c}\right)^2}-e\phi,
\end{equation}
where $\mathbf{P}=\mathbf{p}-e\mathbf{A}/c$ is the canonical momentum and where $\phi$ is the scalar potential. It is useful to employ the co-moving frame variables that move at the wake phase velocity. In the co-moving frame, variables $\xi = v_{\phi}t -x$ and $\tau = x$, the Hamiltonian Eq.~(\ref{eq:hamiltonean}) becomes
\begin{equation}
\label{eq:Hcomoving}
\mathcal{H} = H - v_{\phi}\mathbf{P}_{\|},
\end{equation}
where $\mathbf{P}_{\|}$ corresponds to the canonical momentum in the longitudinal $x$ direction.

In order to determine the constant of motion in multidimensions, we integrate Hamilton's equations written in the co-moving frame. Using the chain rule for the co-moving frame variables
\begin{eqnarray}
\label{eq:chainrule}
\frac{\partial}{\partial x} & = & - \frac{\partial }{\partial \xi}, \\
\frac{\partial}{\partial t} & = & v_{\phi} \frac{\partial }{\partial \xi} + \frac{\partial}{\partial \tau}, \\
\frac{\mathrm{d} \xi}{\mathrm{d} t} & = & v_{\phi}-v_{\|},
\end{eqnarray}
where $v_{\|}=\mathrm{d} x/\mathrm{d} t$ is the longitudinal velocity of an electron. Hamilton's equations can then be written as
\begin{eqnarray}
\label{eq:Heqcomoving}
v_{\phi}\frac{\mathrm{d}P_{\|}}{\mathrm{d}t} & = & -v_{\phi}\frac{\partial  H }{\partial x} =  v_{\phi}\frac{\partial  H }{\partial \xi},  \\
\frac{\mathrm{d} H }{\mathrm{d}t} & = & \frac{\partial H }{\partial t} = v_{\phi}\frac{\partial H }{\partial \xi} + \frac{\partial H }{\partial \tau}.
\end{eqnarray}
Thus, the temporal evolution for $\mathcal{H}$ becomes
\begin{equation}
\label{eq:Htemporal}
\left(v_{\phi}-v_{\|}\right)\frac{\mathrm{d}\mathcal{H}}{\mathrm{d}\xi} = \frac{\partial H}{\partial \tau} = \left[\mathbf{v}\cdot\frac{\partial \mathbf{A}}{\partial \tau}-\frac{\partial \phi}{\partial \tau}\right].
\end{equation}
We note that $\Delta \mathcal{H}$ depends on the initial and final positions only. Thus~\cite{bib:pak_prl_2010,bib:vieira_ppcf_2012,bib:vieira_prl_2011,bib:kalmykov_prl_2009}
\begin{equation}
\label{eq:DeltaH}
\Delta \mathcal{H} = \int \frac{\mathrm{d}\mathcal{H}}{\mathrm{d}t}\mathrm{d} t  = \int \frac{\mathrm{d}\xi}{v_{\phi}-v_{\|}}\frac{\mathrm{d}\mathcal{H}}{\mathrm{d}\xi},
\end{equation}
where the integral on the left-hand side is performed over the particles trajectory~\cite{bib:pak_prl_2010,bib:vieira_ppcf_2012,bib:vieira_prl_2011,bib:kalmykov_prl_2009}. For a non-evolving wakefield, the right-hand side of Eq.~(\ref{eq:DeltaH}) vanishes because $\partial/\partial \tau = 0$. This corresponds to the QSA, where the wakefield does not change during the transit time of a plasma electron. Thus, under the QSA, $\Delta \mathcal{H}=0$, and hence
\begin{eqnarray}
\label{eq:QSAmotion}
\Delta H & =  \Delta \gamma -v_{\phi}\Delta p_{\|} - \left(\Delta \phi - v_{\phi}\Delta A_{\|}\right) \\
& =  \Delta \gamma -v_{\phi}\Delta p_{\|} - \Delta \psi,
\end{eqnarray}
where $\psi = \phi - v_{\phi} A_{\|}$ is called the wake pseudo-potential. Equation~(\ref{eq:QSAmotion}) generalizes the 1D constant of motion given by Eq.~(\ref{eq:constant}) in the presence of magnetic fields (through the presence of the longitudinal vector potential $\Delta A_{\|}$) for particles born with arbitrary initial velocity in regions with arbitrary electric and magnetic fields.

For a particle initially at rest, and initially in a region of vanishing fields, Eq.~(\ref{eq:QSAmotion}) reduces to
\begin{equation}
\label{eq:QSAmotion2}
\gamma\left(1-\beta_{\|}\right) = 1 + \psi.
\end{equation}
As with the 1D case, we can relate the particle velocity to the wake potential $\psi$ and determine the onset of self-injection as a function of $\psi$. When $\beta_{\|}\rightarrow -1$, {i.e.} when particles move backwards at $c$, $\psi\rightarrow \infty$. When $\beta_{\|}\rightarrow 1$, {i.e.} when particles move forward at $c$ and are trapped, $\psi \rightarrow -1$. We can then state that the limit $\psi \rightarrow -1$ is a sufficient, but not necessary, condition for trapping under the QSA.

We will use the constant of motion given by Eq.~(\ref{eq:QSAmotion2}) to derive the equation of motion for plasma electrons that move within the boundary defining the blowout region. The equation of motion is given by the Lorentz force, which can be simplified using Eq.~(\ref{eq:QSAmotion2}). Our calculations now assume $v_{\phi} = c = 1$. Thus, the required time derivative of the equation of motion is
\begin{equation}
\label{eq:timeder}
\frac{\mathrm{d}}{\mathrm{d} t} = \left(1-v_{\phi}\right) \frac{\mathrm{d}}{\mathrm{d} \xi} = \frac{1+\psi}{\gamma} \frac{\mathrm{d}}{\mathrm{d}\xi},
\end{equation}
where we recall that the longitudinal component of the electron velocity $v_{\|}$ is normalized to $c$. As a result, $p_{\perp} = \gamma v_{\perp} =\left(1+\psi\right)\frac{\mathrm{d} r_{\perp}}{\mathrm{d} \xi}$ and
\begin{equation}
\label{eq:dpdt}
\frac{\mathrm{d}p_{\perp}}{\mathrm{d} t} = \frac{1+\psi}{\gamma} \frac{\mathrm{d}}{\mathrm{d}\xi}\left[\left(1+\psi\right)\frac{\mathrm{d}}{\mathrm{d}\xi}\right] r_{\perp},
\end{equation}
where $r_{\perp}$ and $p_{\perp}$ are the transverse radius and momentum, respectively. Equation~(\ref{eq:dpdt}) depends on three quantities, namely $p_{\perp}$, $\gamma$ and $\psi$. It is possible to reduce the number of unknowns by using Eq.~(\ref{eq:QSAmotion2}) to write $\gamma$ as a function of $p_{\perp}$ and $\psi$ as
\begin{equation}
\label{eq:QSAgamma}
\gamma = \frac{1+p_{\perp}^2 + \left(1+\psi\right)^2}{2 \left(1+\psi\right)}.
\end{equation}
Thus, the Lorentz force equation for the radial motion of a plasma electron becomes
\begin{equation}
\label{eq:lorentzqsa}
\frac{2\left(1+\psi\right)^2}{1+\left(1+\psi\right)^2\left(\frac{\mathrm{d}r_{\perp}}{\mathrm{d}\xi}\right)^2 + \left(1+\psi\right)^2} \frac{\mathrm{d}}{\mathrm{d}\xi}\left[\left(1+\psi\right)\frac{\mathrm{d} r_{\perp}}{\mathrm{d}\xi}\right] = F_{\perp},
\end{equation}
where $F_{\perp} = - \left(E_r - v_{\|} B_{\theta}\right)$ is the radial force acting on a plasma electron assuming cylindrical symmetry, with $E_r$ being the radial electric field and $B_{\theta}$ the azimuthal magnetic field. We note that the left-hand side of Eq.~(\ref{eq:lorentzqsa}) depends on $p_{\perp}$, which defines the shape of the bubble, and on $\psi$, which is related to the field structure of the blowout. In order to solve Eq.~(\ref{eq:lorentzqsa}) we need to relate $\psi$ with $p_{\perp}$. This relation can be established through the field structure of the bubble.

The bubble fields can be written as a function of the scalar and vector potentials in the Coulomb Gauge as
\begin{eqnarray}
\label{eq:fields}
E_z & = & \frac{\partial \psi}{\partial \xi} \label{eq:ezblowout}, \\
B_{\theta} &  = & -\frac{\partial A_z}{\partial r} - \frac{\partial A_r}{\partial \xi} \label{eq:bthetablowout}, \\
E_r & = & -\frac{\partial \phi}{\partial r} - \frac{\partial A_r}{\partial \xi} \label{eq:erblowout},
\end{eqnarray}
where scalar and vector potentials can be fully specified once the plasma electron density and currents are known, according to~\cite{bib:lu_msc_thesis}
\begin{eqnarray}
\label{eq:fields-kg}
\frac{1}{r}\frac{\partial}{\partial r}\left(r \frac{\partial A_r}{\partial r}\right) - \frac{A_r}{r^2}  & =  n_\mathrm{e} v_{\perp} & \Leftrightarrow A_r = A_{r0}\left(\xi\right) r \label{eq:potentials_ar}, \\
\frac{1}{r}\frac{\partial}{\partial r}\left(r\frac{\partial A_{\|}}{\partial r}\right) & =  n_\mathrm{b} + n_\mathrm{e} v_{\|} & \Leftrightarrow  A_{\|} = A_{\|0}\left(\xi\right) + \lambda\left(\xi\right) \log\left(r\right) \label{eq:potentials_apar},  \\
\frac{1}{r}\frac{\partial}{\partial r}\left(r\frac{\partial \phi}{\partial r}\right) & =  n_\mathrm{b} + n_\mathrm{e} -1 & \Leftrightarrow \phi = \phi_0 - \frac{r^2}{2} + \lambda\left(\xi\right)\log\left(r\right) \label{eq:potentials_phi}, \\
\frac{1}{r}\frac{\partial}{\partial r}\left(r\frac{\partial \psi}{\partial r}\right) & =  n_\mathrm{e} + n_\mathrm{e} v_{\|} - 1 & \Leftrightarrow \psi = \psi_0\left(\xi\right) -\frac{r^2}{4} \label{eq:potentials:psi}, \\
\frac{1}{r} \frac{\partial}{\partial r} r A_r & =  -\frac{\partial \psi}{\partial \xi} & \Rightarrow A_{r0} = -\frac{1}{2}\frac{\mathrm{d}\psi_0}{\mathrm{d} \xi} \label{eq:potentials_ar2} ,
\end{eqnarray}
where we have included the possibility of describing the effect of a particle beam driver, with $\lambda\left(\xi\right)~=~\int_0^{\infty} r n_\mathrm{b} \mathrm{d} r$ its current profile. Using Eqs.~(\ref{eq:potentials_ar})--(\ref{eq:potentials_ar2}), the right-hand side of Eq.~(\ref{eq:lorentzqsa}) can be rewritten as
\begin{equation}
\label{eq:force}
F_{\perp} = -\frac{r}{2} + \left(1-v_{\|}\right) \frac{\lambda\left(\xi\right)}{r} + \left(1-v_{\|}\right) \frac{\mathrm{d} A_{r0}}{\mathrm{d} \xi} r - \frac{1}{\gamma}\nabla_{\perp} | \frac{a_L}{2}|^2,
\end{equation}
where the first term (on right-hand side of Eq.~(\ref{eq:force})) represents the electrostatic field due to the background plasma ions, the second term is the force exerted on the plasma electrons by a charged particle bunch driver, the third term is due to the radial plasma currents, and the last term is the laser ponderomotive force. By expressing $1-v_{\|} =(1+\psi)/\gamma$ and replacing $\gamma$ by Eq.~(\ref{eq:QSAgamma}), we can rewrite Eq.~(\ref{eq:force}) as
\begin{equation}
\label{eq:lorentzblowout}
\frac{\mathrm{d}}{\mathrm{d}\xi}\left[\left(1+\psi\right)\frac{\mathrm{d} r_\mathrm{b}}{\mathrm{d} \xi}\right] = r_\mathrm{b} \left\{-\frac{1}{4}\left[1+\frac{1}{\left(1+\psi\right)^2}-\left(\frac{\mathrm{d} r_\mathrm{b}}{\mathrm{d} \xi}\right)^2\right]\right\} - \frac{1}{2}\frac{\mathrm{d}^2 \psi_0}{\mathrm{d}\xi^2} + \frac{\lambda\left(\xi\right)}{r_\mathrm{b}^2} - \frac{1}{\left(\psi_0 - \frac{r_\mathrm{b}^2}{4}\right)}\nabla |\frac{a_L}{2}|^2,
\end{equation}
which shows that the trajectory of plasma electrons are fully specified by the pseudo-potential $\psi$. Inversion of Eq.~(\ref{eq:potentials:psi}) reveals that $\psi$ depends on the radial $n_\mathrm{e}(1-v_{\|})$ profile:
\begin{eqnarray}
\label{eq:psiintegralsol}
\psi\left(r,\xi\right) & = & \ln r\int_0^r r^{\prime} \left[n_\mathrm{e}\left(r^{\prime},\xi\right)\left(1-v_{\|}\left(r^{\prime},\xi\right)\right)-1\right] \mathrm{d} r^{\prime}  \\
 & + & \int_r^{\infty} r^{\prime} \ln r^{\prime} \left[n_\mathrm{e}\left(r^{\prime},\xi\right)\left(1-v_{\|}\left(r^{\prime},\xi\right)\right)-1\right] \mathrm{d} r^{\prime}.
\end{eqnarray}

In order to ensure that the fields vanish away from the blowout region, the profile of $n_\mathrm{e}(1-v_{\|})$ is also subject to the following boundary condition:
\begin{equation}
\label{eq:psiboundary}
\int_0^r r^{\prime} \left[n_\mathrm{e}\left(r^{\prime},\xi\right)\left(1-v_{\|}\left(r^{\prime},\xi\right)\right)-1\right] \mathrm{d} r^{\prime}  = 0,
\end{equation}
which states that the source term for $\psi$ $n_\mathrm{e}(1-v_{\|})$ is conserved for each transverse slice.

Equations~(\ref{eq:lorentzblowout}), (\ref{eq:psiintegralsol}) and (\ref{eq:psiboundary}) are general and valid as long as the QSA holds. In order to close our model and to derive the equations that determine the main properties of the blowout regime, we now look for an expression of $\psi$ valid in the blowout regime. In order to determine $\psi$ we then need to find an appropriate model for $n_\mathrm{e}(1-v_{\|})$ which can describe and reproduce the most important features of the blowout regime.

\begin{figure}[ht]
\begin{center}
\includegraphics[width=8 cm]{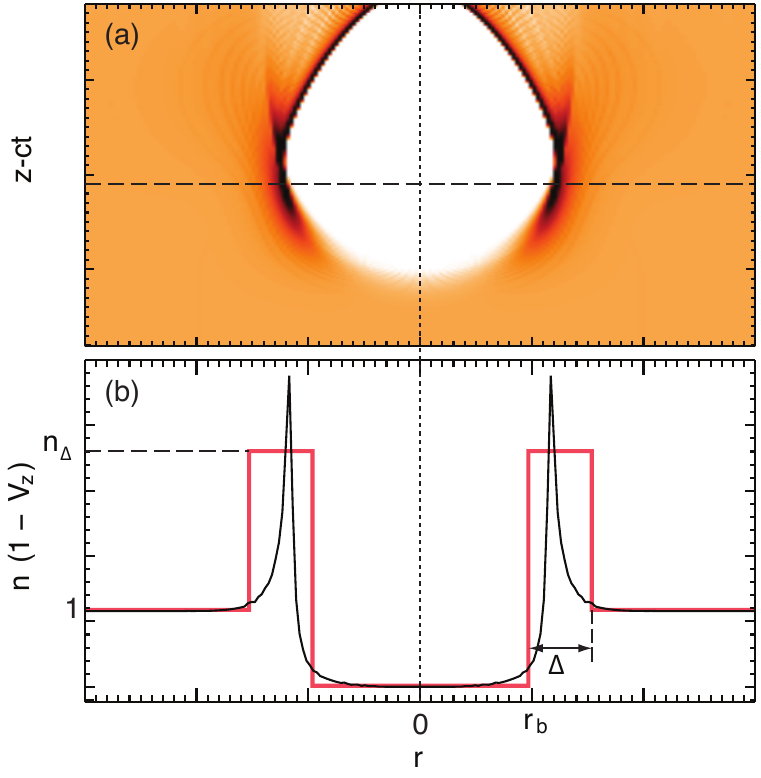}
\caption{PIC simulation result illustrating the blowout regime and the model for the $n_\mathrm{e}(1-v_{\|})$ profile in the blowout: (a) shows $n_\mathrm{e}(1-v_{\|})$ for a slice of the simulation box. The laser driver moves to the bottom of the page; and (b) shows a lineout (solid black line) of $n_\mathrm{e}(1-v_{\|})$ taken at the position of the horizontal dashed line in (a). The solid red line shows the simplified model used to describe the blowout regime.}
\label{fig:blowoutmodel}
\end{center}
\end{figure}

Figure~\ref{fig:blowoutmodel} shows the result of a PIC simulation that illustrates the main properties of the blowout regime that need to be included in a model for $n_\mathrm{e}(1-v_{\|})$ that can accurately describe the blowout regime. Figure~\ref{fig:blowoutmodel} shows that the blowout is characterized by a region empty of plasma electrons, which instead accumulate at the boundary that defines the blowout region. These features are well captured by the simplified model shown by the red line in Fig.~\ref{fig:blowoutmodel}(b), which depends on the blowout radius $r_\mathrm{b}$, on the value of $n_\mathrm{e}(1-v_{\|})=n_{\Delta}$ at $r = r_\mathrm{b}$ and on the thickness $\Delta$ of the electron sheath that defines the blowout.

The boundary condition given by Eq.~(\ref{eq:psiboundary}) yields
\begin{equation}
\label{eq:ndelta}
n_{\Delta}\left(\xi\right) = \frac{r_\mathrm{b}^2}{\left(r_\mathrm{b}+\Delta \right)^2-r_\mathrm{b}^2}.
\end{equation}

Using Eq.~(\ref{eq:psiintegralsol}) and Eq.~(\ref{eq:ndelta}) then yields the following expression for $\psi$ as a function of $r_\mathrm{b}$ and $\alpha = \Delta/r_\mathrm{b}$:
\begin{equation}
\label{eq:psisolution}
\psi\left[r_\mathrm{b}\left(\xi\right)\right] = \frac{r_\mathrm{b}^2}{4} \left(\frac{\left(1+\alpha\right)^2\ln\left(1+\alpha\right)^2}{\left(1+\alpha\right)^2}-1\right) \equiv \frac{r_\mathrm{b}^2}{4} \beta.
\end{equation}
Equation~(\ref{eq:psisolution}) is a general expression that specifies $\psi$ for given a $r_\mathrm{b}$ and $\alpha$. The plasma response in the blowout regime is always non-linear, but, depending on the maximum value of $r_\mathrm{b}$, it may be relativistic or non-relativistic. When $k_\mathrm{p} r_\mathrm{b}\ll 1$, the plasma response is non-relativistic. In this limit, the ratio of the blowout radius to the width of the thin electron sheath is much smaller than unity, or $\alpha = \Delta/r_\mathrm{b} \gg 1$. In the ultra-relativistic blowout regime, $k_\mathrm{p} r_\mathrm{b} \gg 1$ and $\alpha =  \Delta/r_\mathrm{b}  \ll 1$, i.e. the width of the electron layer that defines the blowout is much smaller than the blowout radius. We can use Eq.~(\ref{eq:psisolution}) to find $\psi$ in these two limiting scenarios. Hence, in the non-relativistic blowout regime, Eq.~(\ref{eq:psisolution}) reduces to
\begin{equation}
\label{eq:psinonrelativistic}
\psi\left(r,\xi\right) \simeq  \frac{r_\mathrm{b}^2(\xi)}{4}\log\left(\frac{1}{r_\mathrm{b}}\right) - \frac{r^2}{4},
\end{equation}
and in the ultra-relativistic blowout regime, Eq.~(\ref{eq:psisolution}) becomes
\begin{equation}
\label{eq:psiultrarelativistic}
\psi\left(r,\xi\right) \simeq \left(1+\alpha\right) \frac{r_\mathrm{b}^2(\xi)}{4} - \frac{r^2}{4}.
\end{equation}

\begin{figure}[ht]
\begin{center}
\includegraphics[width=8 cm]{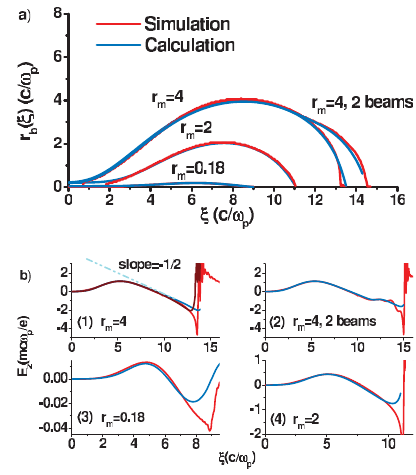}
\caption{(Picture taken from Ref.~\cite{bib:lu_prl_2006}) Comparison between the blowout theory (blue lines) and PIC simulations (red lines): (a) trajectory of the blowout radius $r_\mathrm{b}(\xi)$ for different maximum blowout radius $r_\mathrm{m}$, from the non-relativistic blowout regime ($k_\mathrm{p} r_\mathrm{m}=0.18$) and strongly relativistic blowout regime ($k_\mathrm{p} r_\mathrm{m} = 4$); and (b) shows the corresponding on-axis accelerating fields for all the cases in (a). The driver propagates from right to left, in the direction of negating $\xi$.}
\label{fig:blowoutcomparison}
\end{center}
\end{figure}

In order to obtain a general description of the blowout regime, valid in the relativistic and non-relativistic regimes, we insert the expression for $\psi$ given by Eq.~(\ref{eq:psisolution}) into the Lorentz force equation given by Eq.~(\ref{eq:lorentzblowout}). This results in a non-linear differential equation for the motion of the inner most electron in the blowout $r_\mathrm{b}(\xi)$~\cite{bib:lu_prl_2006}
\begin{equation}
\label{eq:blowoutequation}
A(r_\mathrm{b}) \frac{\mathrm{d}^2 r_\mathrm{b}}{\mathrm{d} \xi^2} + B(r_\mathrm{b}) r_\mathrm{b} \left(\frac{\mathrm{d} r_\mathrm{b}}{\mathrm{d} \xi}\right)^2 + C(r_\mathrm{b}) r_\mathrm{b} = \frac{\lambda(\xi)}{r_\mathrm{b}} - \frac{1}{4}\frac{\mathrm{d}|a|^2}{\mathrm{d}r} \frac{1}{\left(1+\beta r_\mathrm{b}^2/4\right)^2},
\end{equation}
where we have assumed that $\Delta$ does not depend on $\xi$, and where
\begin{eqnarray}
\label{eq:blowoutdefinitions}
A(r_\mathrm{b}) & = & 1 + \left(\frac{1}{4}+\frac{\beta}{2}+\frac{1}{8}r_\mathrm{b} \frac{\mathrm{d}\beta}{\mathrm{d}r_\mathrm{b}}\right) r_\mathrm{b}^2, \\
B(r_\mathrm{b}) & = & \frac{1}{2} + \frac{3}{4}\beta + \frac{3}{4}r_\mathrm{b} \frac{\mathrm{d}\beta}{\mathrm{d} r_\mathrm{b}} + \frac{1}{8} r_\mathrm{b}^2 \frac{\mathrm{d}^2 \beta}{\mathrm{d} r_\mathrm{b}^2}, \\
C(r_\mathrm{b}) & = & \frac{1}{4}\left(1+\frac{1+|a|^2/2}{1+\beta r_\mathrm{b}^2/4}\right).
\end{eqnarray}
In the ultra-relativistic blowout regime ($k_\mathrm{p} r_\mathrm{b} \gg 1$), the assumption that $\partial_{\xi} \Delta \simeq 0$ breaks at the back of the bubble where $\Delta \ll r_\mathrm{b}$. Numerical solutions to Eqs.~(\ref{eq:blowoutequation}) and (\ref{eq:blowoutdefinitions}) are in very good agreement with full PIC simulations for a wide range of conditions from weakly- to ultra-relativistic blowouts. The agreement is nearly perfect except for the region at the back of the bubble where $\Delta \simeq r_\mathrm{b}$. Figure~\ref{fig:blowoutcomparison}(a), which compares Eq.~(\ref{eq:blowoutequation}) with PIC simulations, shows excellent agreement for almost the entire blowout region, except at the back of the bubble. Figure~\ref{fig:blowoutcomparison} considered an electron beam driver to excite the blowout. In the laser driven case, the blowout region is not as well defined. Still, comparisons between theory and simulations are very good when $W_0 \simeq r_\mathrm{m}$, where $r_\mathrm{m}$ is the maximum blowout radius.

It is possible to simplify Eq.~(\ref{eq:blowoutequation}) in order to predict key wakefield properties in current plasma acceleration experiments, which operate in the ultra-relativistic regime. The shape of the blowout region can be predicted in this extreme regime by assuming $\alpha \ll 1$, for which Eq.~(\ref{eq:blowoutequation}) simplifies to
\begin{equation}
\label{eq:blowoutrelativistic}
r_\mathrm{b} \frac{\mathrm{d}^2 r_\mathrm{b}}{\mathrm{d}\xi^2} + 2 \left(\frac{\mathrm{d} r_\mathrm{b}}{\mathrm{d}\xi}\right)^2+1 =  \frac{4 \lambda(\xi)}{r_\mathrm{b}} - \frac{\mathrm{d}|a|^2}{\mathrm{d}r} \frac{1}{\left(1+\beta r_\mathrm{b}^2/4\right)^2}.
\end{equation}
At the back of the driver where the right-hand side of Eq.~(\ref{eq:blowoutrelativistic}) vanishes, Eq.~(\ref{eq:blowoutrelativistic}) becomes very similar to the equation of a sphere, which is given by
\begin{equation}
\label{eq:sphere}
r_\mathrm{b} \frac{\mathrm{d}^2 r_\mathrm{b}}{\mathrm{d}\xi^2} + \left(\frac{\mathrm{d} r_\mathrm{b}}{\mathrm{d}\xi}\right)^2+1 = 0.
\end{equation}
The main difference between Eqs.~(\ref{eq:blowoutrelativistic}) and (\ref{eq:sphere}) is the multiplication factor in the second term on the left-hand side of both equations. The additional factor of two in Eq.~(\ref{eq:blowoutrelativistic}) leads to a stronger bending of the blowout radius at the back of the bubble when compared with a sphere. However, for most of the wakefield, the blowout resembles a sphere.

Having determined $r_\mathrm{b}$, and hence $\psi$, it is now possible to derive the full field structure of the blowout region. The accelerating field is given by Eq.~(\ref{eq:ezblowout}), which reads $E_z = \partial \psi /\partial \xi$, and the focusing field acting on a relativistic particle traveling at $c$ is $W_{\perp}=E_r - B_{\theta} = -\partial\psi/\partial r$. In ultra-relativistic regimes, $E_z$ can be evaluated by integrating Eq.~(\ref{eq:blowoutrelativistic}) near the region where the blowout radius is maximum, and then inserting the resulting expression for $r_\mathrm{b}$ into Eq.~(\ref{eq:psiultrarelativistic}). Using Eq.~(\ref{eq:ezblowout}) gives
\begin{equation}
\label{eq:ezblowout2}
E_z = \frac{1}{2}\frac{\mathrm{d} r_\mathrm{b}}{\mathrm{d} \xi} \simeq \frac{\xi}{2}.
\end{equation}
The maximum accelerating gradient is thus given by $E_z \simeq r_\mathrm{b}/2$, and we recover the result from the phenomenological theory for the blowout regime. Figure~\ref{fig:blowoutcomparison}(b) compares the theoretical predictions for $E_z$ with simulation results from the non-relativistic to the ultra-relativistic blowout regime, showing very good agreement, except at the back of the bubble where the blowout approaches the axis. We note that, unlike in the linear regime, the accelerating electric field in the bubble is independent of $r$.

The focusing force can be determined in a similar manner yielding:
\begin{equation}
\label{eq:focusingblowout}
E_r - B_{\theta} = \frac{r}{2},
\end{equation}
which is also in excellent agreement with simulation results.

\section{Beam loading}
\label{sec:beamloading}

In the previous section we demonstrated that the blowout regime provides linear focusing and accelerating fields. Linear focusing fields are critical to preserve the emittance of accelerated beams during their acceleration. Matching conditions, relating the initial beam emittance to the linear focusing force in a plasma are well known (see, for instance, \cite{bib:clayton_prl_2002} and references therein). Linear accelerating fields suggest a way to tailor the currents of accelerated particle bunches ensuring constant accelerating fields for all bunch particles. This is critical for the acceleration of particles with no energy spread growth~\cite{bib:tzoufras_prl_2008}.

In order to investigate beam loading in the strongly non-linear blowout regime we integrate Eq.~(\ref{eq:blowoutrelativistic}):
\begin{equation}
\label{eq:ezbeamloading}
E_z = \frac{1}{2} r_\mathrm{b} \frac{\mathrm{d} r_\mathrm{b}}{\mathrm{d}\xi}=-\frac{r_\mathrm{b}}{2\sqrt{2}}\sqrt{\frac{16\int l(\xi)\xi\mathrm{d}\xi+C}{r_\mathrm{b}^4}-1},
\end{equation}
where $l$ is the current density of the electron beam injected into the plasma wave. Equation~(\ref{eq:ezbeamloading}) gives the accelerating electric field in the blowout regime as a function of the position in the bubble $\xi$ and for an arbitrary beam loading current profile. It is interesting to note that the beam loading depends on the integrated current profile of the beam, but not on the particular transverse shape of the bunch. Thus, the accelerating field remains unchanged even if the bunch evolves transversely.

\begin{figure}[ht]
\begin{center}
\includegraphics[width=8.5 cm]{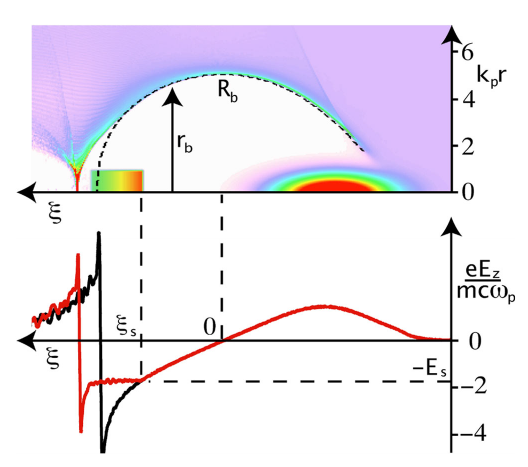}
\caption{(Picture taken from Ref.~\cite{bib:tzoufras_prl_2008}) Top plot shows plasma electron density in the blowout, superimposed with a particle beam driver and a witness beam at the back of the bubble. The driver propagates from left to right. The blowout radius $r_\mathrm{b}$ and maximum blowout radius $R_\mathrm{b}$ are also indicated. The dashed line in the top plot shows the shape of the bubble in an unloaded scenario where the witness bunch is absent. The initial location of the witness bunch ($\xi_s$) and corresponding accelerating field $E_s$ is also indicated in the bottom plot. The red (black) lines in the bottom plot show the accelerating fields with (without) the witness electron bunch.}
\label{fig:beamloading}
\end{center}
\end{figure}

The integral in Eq.~(\ref{eq:ezbeamloading}) can be calculated analytically for the case of a trapezoidal bunch. The optimal beam loading current for a trapezoidal bunch, ensuring constant accelerating electric fields along the bunch is
\begin{equation}
\label{eq:optimalloading}
l(\xi) = \sqrt{E_s^4+\frac{R_\mathrm{b}^4}{16}} - E_s\left(\xi - \xi_s\right),
\end{equation}
where the meaning of $E_s$ and $\xi_s$ is shown in Fig.~\ref{fig:beamloading}. If the beam current profile satisfies Eq.~(\ref{eq:optimalloading}), as in the case of Fig.~\ref{fig:beamloading}, then all bunch particles accelerate with similar acceleration gradients, hence minimizing energy spread variations throughout the propagation. The trailing electron bunch will inevitably be subject to different accelerating fields as it dephases in the plasma wave in laser wakefield accelerators. Thus, growth of energy spread is expected in laser wakefield accelerators. This may be compensated by phase-space rotation near dephasing~\cite{bib:tsung_prl_2004}.

The maximum charge that can be loaded into the wakefield can be determined by assuming that the back of the bunch coincides with $r_\mathrm{b} = 0$. Using Eq.~(\ref{eq:optimalloading}), the maximum charge is thus given by
\begin{equation}
\label{eq:trappedq}
Q_{s} = \frac{\pi}{16}\frac{R_\mathrm{b}^4}{E_s},
\end{equation}
where $R_\mathrm{b} = r_\mathrm{b}(\xi = 0)$ ($\xi = 0$ is at the centre of the bubble, cf. Fig.~\ref{fig:beamloading}). Equation~(\ref{eq:trappedq}) shows that smaller $E_s$ leads to higher charges because the beam can be made longer. However, the maximum energy per particle is also lower for smaller $E_s$. This illustrates the trade-off between the maximum energy gain and the maximum number of accelerated particles. The accelerator efficiency is the ratio between the absorbed energy and the total wakefield energy. This can be expressed as
\begin{equation}
\label{eq:etablowout}
\eta = \frac{\tilde{Q}_s}{Q_s},
\end{equation}
where $\tilde{Q}_s$ is the charge of a trapezoidal bunch described by Eq.~(\ref{eq:optimalloading}).

\section{Laser wakefield acceleration driven by structured light}
\label{sec:structuredlight}

Structured laser pulses have non-trivial spatiotemporal phase structures, similar to those appearing when multiple plane waves interfere. The interference of three plane waves, for example, can lead to the orbital angular momentum (OAM) of light~\cite{bib:allen_pra_1992}. The OAM  has been studied over the last 30 years because of its usefulness in a broad range of scientific and technological applications. Twisted lasers with OAM can be used to control the poloidal motion of nano-particles in twisted optical tweezers~\cite{bib:padgett_pt_2004} and obtain super-resolution microscopic images~\cite{bib:hell_ol_1994}. These lasers also promise a dramatic increase of the speed of optical communications~\cite{bib:wang_natphoton_2012,bib:zhang_science_2020}, and may provide pathways to detect spinning blackholes~\cite{bib:tamburini_natphys_2011}.

The exotic phase structure of OAM laser pulses is directly responsible for these scientific and technological advances. OAM lasers have a helical wave-front structure with a phase-singularity on axis. The intensity needs to vanish at the phase-singularity because the phase is ill defined. As a result, these modes have a doughnut shaped intensity profile. Mathematically, these lasers correspond to Laguerre-Gaussian modes, which are the higher order solutions of the electromagnetic wave equation written in cylindrical coordinates.

Lasers with orbital angular momentum are often produced with conventional optical elements, such as computer generated holograms or spiral phase plates. Although such techniques work best at intensities below damage thresholds, some of them can be applied at higher intensities too. Producing OAM lasers at very high intensity is important because this may unlock completely new territory in plasma based accelerators. Before further elaborating on the topic, it is useful to first describe on-going research towards the generation of intense lasers with OAM.

Since plasmas have no damage thresholds, they can be used to generate and to amplify OAM lasers to very high intensities, well above ionisation thresholds, for example via stimulated Raman amplification~\cite{bib:vieira_natcomms_2016}. In Raman amplification, a long and energetic laser probe pulse beats with a short, low energy seed pulse in plasma. The beating pattern creates a Langmuir plasma wave that transfers the energy from the probe pulse to the seed pulse. Depending on the OAM structure of the probe pulse, this process can either maintain the OAM level of the seed, or create high OAM harmonics while keeping the laser pulse frequency constant~\cite{bib:vieira_prl_2016}. If maintained over ion time scales, Langmuir waves created in stimulated Raman scattering can also create a periodic refractive index structure in the plasma. These structures are akin to photonic crystals, which are now standard in the fields of optics and photonics. Photonic crystals display unprecedented properties to control light. Plasma based photonic crystals may thus bring these optical elements into the ultra-high intensity realm~\cite{bib:lehmann_prl_2016}, and can be used to produce intense lasers with OAM~\cite{bib:lehmann_pre_2019}.

In Raman amplification, the plasma acts as a transmissive optical element. Reflective optical elements can also efficiently imprint OAM into high intensity laser pulses. For example, interfering two laser beams at an angle at a solid surface can create a plasma based hologram. Under certain circumstances, such hologram can display a fork-structure, required to imprint OAM on an initially Gaussian laser pulse after reflection~\cite{bib:leblanc_nphys_2017}. An on-axis plasma based spiral phase mirror may be used to imprint OAM into a previously focused Gaussian laser pulse~\cite{bib:shi_prl_2014}. Furthermore, conventional reflective optical elements have already been shown to provide a robust pathway to produce OAM lasers at very high intensity. Off-Axis Spiral Phase Mirrors are a promising example, which was used to produce OAM laser pulses at ultra-high intensity experimentally~\cite{bib:longman_ol_2020}. 

The spatiotemporal profile of structured lasers, at ultra-high intensities, enables to control the plasma dynamics in ways that are not directly accessible to Gaussian laser pulse drivers. As a result of such control, structured light can open new routes not only to address on-going challenges in plasma acceleration but also to expand plasma accelerators beyond current possibilities. An example is positron acceleration in the nonlinear regime, which is not effective in the conventional blowout because the blowout region defocuses positrons. Unlike Gaussian drivers, OAM laser pulses can excite doughnut shaped, nonlinear plasma wakefields~\cite{bib:mendonca_pop_2014}. The doughnut wakefield shape closely resembles the intensity profile of Laguerre-Gaussian modes, as plasma electrons respond directly to the laser ponderomotive force.

\begin{figure}[ht]
\begin{center}
\includegraphics[width=8 cm]{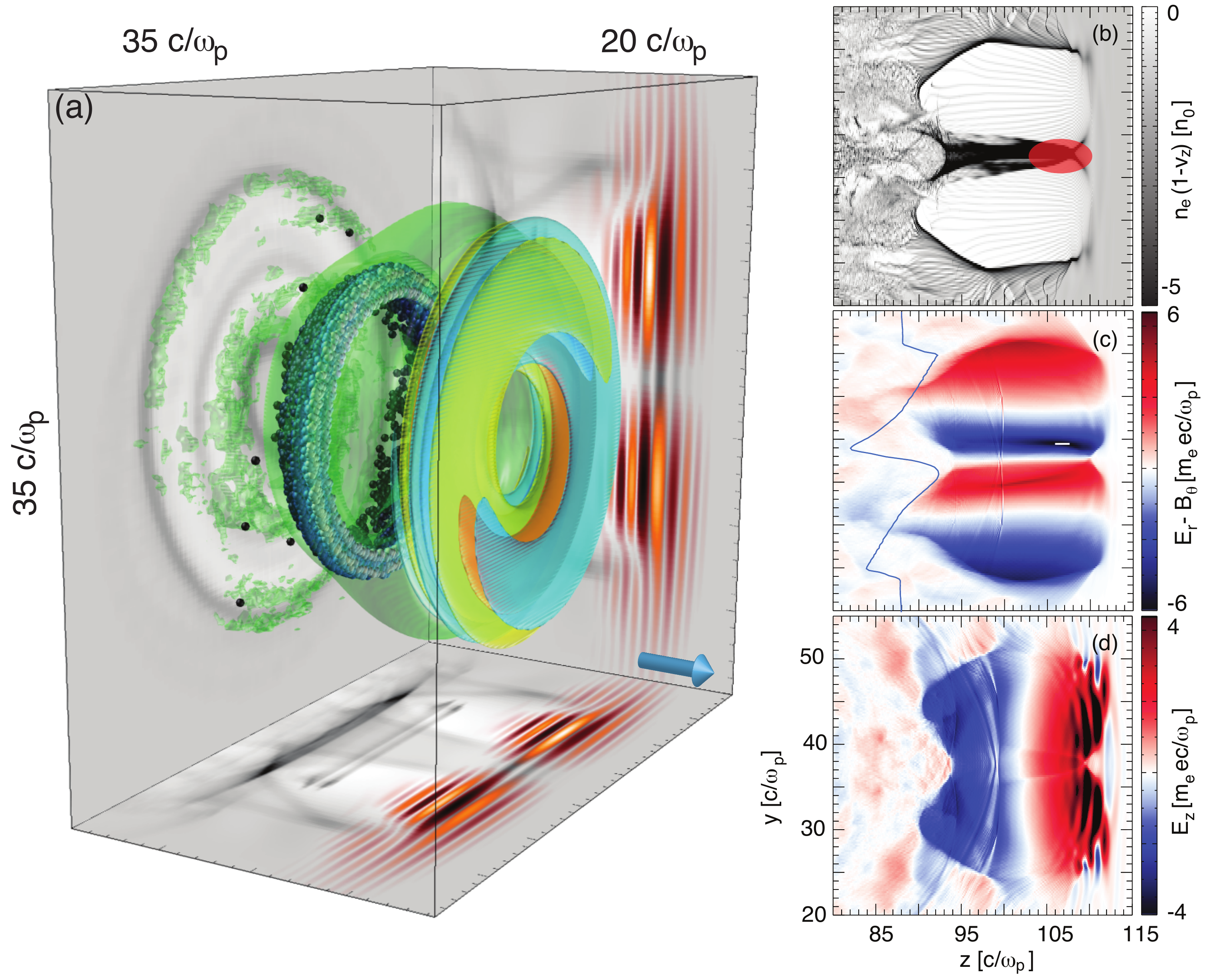}
\caption{PIC simulation results showing structures leading to positron acceleration in the strongly non-linear blowout regime using a Laguerre-Gaussian beam driver with orbital angular momentum. (a) shows plasma density isosurfaces (green) illustrating the generation of a doughnut bubble. Plasma density projections are in grey, and laser pulse electric field projections in orange. The laser propagates in the direction of the blue arrow. Blue spheres represent a self-injected electron bunch ring. (b) shows the plasma density structures taken from a central slice of the simulation box. The location for positron focusing and accelerating fields is shown by the red shape. (c)-(d) shows the focusing and accelerating field structures taken from a central slice of the simulation box. The blue line in (c) shows a lineout of the focusing wakefields.}
\label{fig:figure11}
\end{center}
\end{figure}

Figure~\ref{fig:figure11}a illustrates the spatial structure of a doughnut shaped plasma wave. The trajectory of an inner and outer plasma electron thin sheath define the hollow region of the doughnut wakefield. The onset of positron focusing and acceleration occurs when the laser ponderomotive force pushes the inner electron sheath to the axis, as seen in Fig.~\ref{fig:figure11}b. The on-axis electron sheath carries a charge density that is much higher than the background plasma ion density. The ensuing radial electric field therefore provides a positron focusing force on-axis, shown in Fig.~\ref{fig:figure11}c. Additionally, longitudinal electric fields, depicted in Fig.~\ref{fig:figure11}d, which are defined by the trajectory of the outer doughnut wakefield electron sheath can accelerate positrons at the front of the bubble~\cite{bib:vieira_prl_2014}. In general, higher order solutions of the electromagnetic wave-equation provide laser pulse drivers that effectively control the focusing fields in plasma wakefields. Multiple combinations of higher order Hermite-Gaussian beams, for example, could then effectively tailor positron focusing and acceleration in nonlinear plasma wakefields~\cite{bib:yu_pop_2014}. Particle bunches with ring-shaped current profiles could provide similar structures for positron acceleration~\cite{bib:jain_arxiv_2014}.

In addition to positron acceleration, doughnut shaped wakefields driven by Laguerre-Gaussian laser pulses can also be used to generate and accelerate ring-shaped electron bunches to relativistic energies~\cite{bib:zhang_jap_2016}, via ionisation injection techniques~\cite{bib:zhang_pop_2016}. Electron bunches with these profiles could be used to accelerate and collimate positively charged particles. Furthermore, intense lasers with OAM could be used to control the topological properties of accelerated bunches~\cite{bib:ju_njp_2018}.

Laser pulses with OAM can excite wakefields with new topologies and provide control over previously inaccessible phase-space features of accelerated particles. Twisted plasma waves, that carry orbital angular momentum because of a helical phase structure~\cite{bib:blackmann_pre_2019}, are well suited to control the angular momentum degrees of freedom of relativistic charged particle bunches. Intense lasers with helical intensity profiles, also denoted as spatiotemporal Light Springs~\cite{bib:pariente_ol_2015}, can excite twisted plasma waves. The helical intensity profile of a light spring results from a spatio-spectral correlation between the laser OAM and frequency. For example, beating two Laguerre Gaussian modes with different frequencies and OAM levels creates such a helical intensity profile. The generation of spatiotemporal Light Springs could be accomplished using spiral phase plates. Thick spiral phase plates will induce a group velocity dispersion that varies azimuthally, thus imprinting a helical intensity profile on an initially, ultra-short Gaussian laser pulse.

\begin{figure}
\centering\includegraphics[width=8 cm]{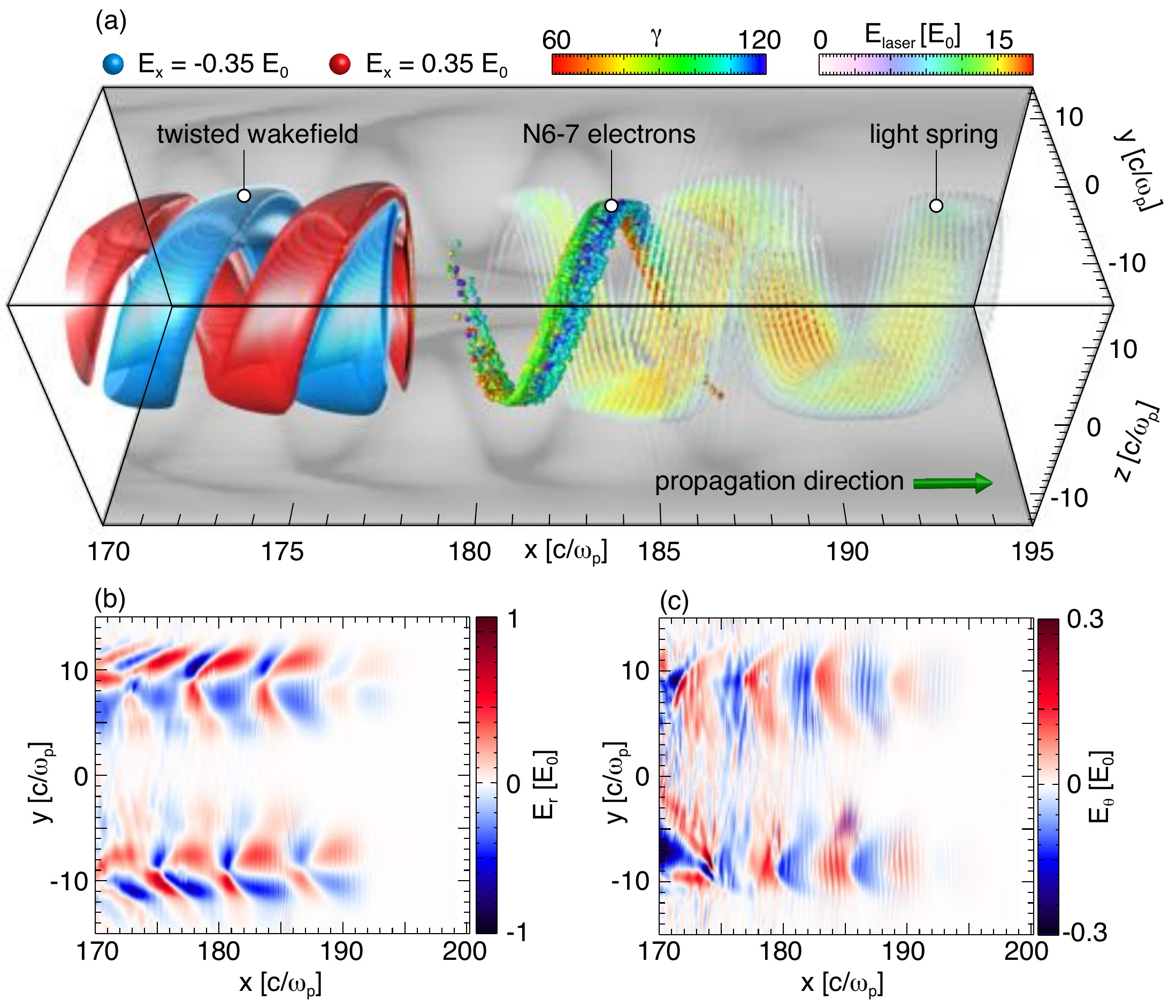}
\caption{Twisted plasma wakefields driven by a light spring moving in a preformed plasma doped with Nitrogen. (a) (right) Rainbow colors display the electric field of the light spring. (left) blue-red isosurfaces show the twisted longitudinal electric field structure of the wake excited by this LS in the underdense plasma. These surfaces are not displayed for $x \geq 178$ to avoid hiding the other plots. (middle) Spheres in rainbow colors correspond to ionization injected electrons from the inner ($6^{th}-7^{th}$) shells of Nitrogen. (b)-(c) slices of the corresponding radial and azimuthal electric fields in the plasma. Field values are normalised to the cold wavebreaking limit $E_0=m_e c \omega_p/e$.}
\label{fig:figure12}
\end{figure}

Efficient wakefield excitation occurs when the pitch of the helical intensity pattern is equal to the plasma frequency. Figure~\ref{fig:figure12}a illustrates the spatial structure of a twisted plasma wakefield, in the nonlinear regime, driven by a spatiotemporal Light Spring. In addition to longitudinal and radial (Fig.~\ref{fig:figure12}b) components, twisted plasma waves also carry an azimuthal wakefield component that can impart angular momentum to trapped particles, which is shown in Fig.~\ref{fig:figure12}c. It follows from the twisted wakefield structure that trapped particles can only travel with preset azimuthal velocities. This corresponds to a quantisation of their angular momentum. Additionally, the current profile of trapped particles consists of one or more inter-twined helixes, which follow the twisted wakefield structure OAM. Figure~\ref{fig:figure12}a provides an illustration of a relativistic electron bunch consisting of a single helix. Hence, twisted wakefields can generate classical electron bunches with orbital angular momentum~\cite{bib:vieira_prl_2018}. Such bunches are interesting because they may relax the standard criteria for superradiant spontaneous undulator radiation~\cite{bib:vieira_2020}, and may thus open new routes to produce temporally coherent radiation in plasma accelerators. These bunches may also be useful for the multiple communities that use charged particle bunches to explore magnetic properties of matter~\cite{bib:harris_nphys_2015}.

The OAM is one of the many examples of advanced spatiotemporal couplings that promises to unlock unprecedented degrees of freedom to control laser-plasma accelerators. Other types of spatiotemporal couplings may play a crucial role in overcoming current limits in laser-plasma accelerators, namely the dephasing length limit. The dephasing length is the distance travelled by accelerated particles before they outrun the accelerating phase of the wakefield. It is useful to estimate the dephasing length by determining the distance travelled by a relativistic particle relative to the plasma wakefield. This distance is $\Delta x = (v_p-v_{\phi}) L/c$, where $v_p \simeq c $ is the particle velocity, $v_{\phi}$ is the wakefield phase-velocity, and $L$ is the distance travelled in the laboratory frame. In the non-linear regime, dephasing occurs when $\Delta x \simeq \lambda_p/2$. The corresponding dephasing length is thus $L_d \simeq \lambda_p/[2 (1-v_{\phi}/c)]\simeq \lambda_p \gamma_{\phi}^2$, where $\gamma_{\phi}^2=1/(1-v_{\phi}^2/c^2)$ is the wakefield relativistic factor. According to the latter expression for $L_d$, dephasing can be ignored in plasma wakefield accelerators where  $\gamma_{\phi} \simeq 10^4-10^5$. In contrast, dephasing limits the acceleration distance in laser wakefield accelerators, because $\gamma_{\phi}\simeq 10-100$ in current experiments.

The discussion above shows that dephasingless acceleration is a desirable feature in laser-plasma accelerators. In the standard paradigm, the plasma density defines the wakefield phase velocity. Thus, the output energy of laser plasma accelerators is intimately related to the plasma density: higher energies require plasma wakefields with higher $v_{\phi}$, which, in turn, can only operate at lower plasma densities, which in turn support lower acceleration gradients and longer acceleration distances. Obtaining independent control over the wakefield phase velocity and the plasma density can thus enable dephasingless acceleration regimes at any plasma density. Dephasingless acceleration in laser driven wakefields requires plasma wakefields travelling at $c$. This could be achieved by relying on specific spatiotemporal couplings on the laser driver. The flying/sliding focus technique~\cite{bib:marie_optica_2017,bib:froula_nphoton_2018}, an appropriate use of an axi-parabola~\cite{bib:palastro_prl_2020}, traveling wave electron acceleration~\cite{bib:debus_prx_2019} are three examples of how advanced spatiotemporal couplings could be used to overcome the dephasing acceleration limit.

\section{Conclusions}
\label{sec:conclusions}

In this report we have outlined wakefield excitation models using various approximations, valid in one- and multidimensional scenarios. Using these models, we have shown how the field structure of plasma waves in the linear and non-linear regimes can be determined, and we have derived a set of scaling laws for the maximum energies and charges that can be achieved in plasma-based accelerators. We have shown that the blowout regime has the potential to accelerate high-charge, high-quality electron bunches to high energies. Laser wakefield acceleration experiments in the blowout regime have demonstrated multi-GeV electron accelerations~\cite{bib:leemans_prl_2014,bib:litos_nature_2014}, and future experiments in the field promise to take the technology even closer to applications. We have also briefly reviewed the role of structured laser drivers to overcome and expand laser wakefield accelerators beyond current limits of both conventional and advanced acceleration techniques.

\section*{Acknowledgements}

We acknowledge PRACE for access to resources on SuperMUC (Leibniz Research Center). This work was supported by the European Research Council (ERC-2015-AdG Grant695088, by the EU Accelerator Research and Innovationfor European Science and Society (EU ARIES) GrantAgreement No. 730871 (H2020-INFRAIA-2016-1) and FCT  (Portugal) GrantNo. SFRH/IF/01635/2015.


\begin{thebibliography}{99}
\raggedright
\bibitem{bib:tajima_prl_1979} T. Tajima and J.M. Dawson, \emph{Phys. Rev. Lett.}~\textbf{43} (4) (1979) 267--270. http://dx.doi.org/10.1103/PhysRevLett.43.267
\bibitem{bib:eli} ELI beam lines facility, http://www.eli-beams.eu; ELI Nuclear Physics, http://www.eli-np.ro; ELI Atto-second, http://www.eli-hu.hu.
\bibitem{bib:silva_pre_1999} L.O. Silva \emph{et al.}, \emph{Phys. Rev.} E \textbf{59}(2) (1999) 2273. http://dx.doi.org/10.1103/PhysRevE.59.2273
\bibitem{bib:esarey_ieee_1996} E. Esarey \emph{et al.}, \emph{IEEE Trans. Plasma Sci.}~\textbf{24}(2) (1996) 252. http://dx.doi.org/10.1109/27.509991
\bibitem{bib:geddes_nature_2004} C.G.R. Geddes \emph{et al.}, \emph{Nature}~\textbf{431}  (2004) 538. http://dx.doi.org/10.1038/nature02900
\bibitem{bib:faure_nature_2004} J. Faure \emph{et al.}, \emph{Nature}~\textbf{431}  (2004)  541. http://dx.doi.org/10.1038/nature02963
\bibitem{bib:mangles_nature_2004} S.P.D. Mangles \emph{et al.}, \emph{Nature}~\textbf{431}  (2004) 535. http://dx.doi.org/10.1038/nature02939
\bibitem{bib:dawson_pr_1959} John M. Dawson, \emph{Phys. Rev.}~\textbf{113}(2)  (1959) 383. http://dx.doi.org/10.1103/PhysRev.113.383
\bibitem{bib:leemans_prl_2014} W.P. Leemans \emph{et al.}, \emph{Phys. Rev. Lett.}~\textbf{113}(24) (2014) 245002. http://dx.doi.org/10.1103/PhysRevLett.113.245002
\bibitem{bib:litos_nature_2014} M. Litos \emph{et al.},  \emph{Nature}~\textbf{515}  (2014) 92. http://dx.doi.org/10.1038/nature13882
\bibitem{bib:osiris} R.A. Fonseca \emph{et al.}, OSIRIS: A Three-Dimensional, Fully Relativistic Particle in Cell Code for Modeling Plasma Based Accelerators, \emph{Lecture Notes in Computer Science} Vol. 2331/2002 (Springer, Berlin, Heidelberg 2002), pp. 342--351. http://dx.doi.org/10.1007/3-540-47789-6$\_$36
\bibitem{bib:quickpic} C. Huang \emph{et al.},  \emph{J. Comp. Phys.}~\textbf{217}(2) (2006) 658. http://dx.doi.org/10.1016/j.jcp.2006.01.039
\bibitem{bib:fonseca_ppcf_2013} R.A. Fonseca \emph{et al.}, \emph{Plasma Phys. Control. Fusion}~\textbf{55}(12) (2013) 124001. http://dx.doi.org/10.1088/0741-3335/55/12/124011
\bibitem{bib:master1} P. Gibbon, \emph{Short Pulse Laser Interactions with Matter: An Introduction} (Imperial College Press, Published by World Scientific Publishing Company, London, 2005). http://dx.doi.org/10.1142/p116
\bibitem{bib:master2} W.B. Mori \emph{et al}, {Private communication}.
\bibitem{bib:sprangle_prl_1990} P. Sprangle \emph{et al.}, \emph{Phys. Rev. Lett.}~\textbf{64}(17) (1990) 2011. http://dx.doi.org/10.1103/PhysRevLett.64.2011
\bibitem{bib:russian_qsa}V.I. Berezhiani and I.G. Murusidze, \emph{Phys. Lett.} A~\textbf{148}(6--7) (1990) 338. http://dx.doi.org/10.1016/0375-9601(90)90813-4
\bibitem{bib:wake} Thomas M. Antonsen and P. Mora, \emph{Phys. Plasmas}~\textbf{4}(1) (1997) 217. http://dx.doi.org/10.1063/1.872134
\bibitem{bib:desy_qsa} T Mehrling \emph{et al.},  \emph{Plasma Phys. Control. Fusion}~\textbf{56}(8) (2014) 084012. http://dx.doi.org/10.1088/0741-3335/56/8/084012
\bibitem{bib:katsouleas_bemloading} T. Katsouleas \emph{et al}, \emph{Particle Accelerators} \textbf{22}  (1987) 81.
\bibitem{bib:lu_prl_2006} W. Lu \emph{et al.}, \emph{Phys. Rev. Lett.}~\textbf{96}(16) (2006) 165002. http://dx.doi.org/10.1103/PhysRevLett.96.165002
\bibitem{bib:lu_msc_thesis} W. Lu, Master Thesis, University of California Los Angeles, 2004.
\bibitem{bib:pukhov_apl_2002} A. Pukhov and J. Meyer-Ter-Vehn, \emph{Appl. Phys.} B~\textbf{74} (4--5) (2002) 355. http://dx.doi.org/10.1007/s003400200795
\bibitem{bib:lee_pre_2001} S. Lee \emph{et al.}, \emph{Phys. Rev.} E~\textbf{64}(4) (2001) 045501(R). http://dx.doi.org/10.1103/PhysRevE.64.045501
\bibitem{bib:lu_prstab_2007} W. Lu \emph{et al.}, \emph{Phys. Rev.} ST-AB~\textbf{10}(6) (2007) 061301. http://dx.doi.org/10.1103/PhysRevSTAB.10.061301
\bibitem{bib:decker_pop_1996} C.D. Decker \emph{et al.}, \emph{Phys. Plasmas}~\textbf{3}(4) (1996) 1360. http://dx.doi.org/10.1063/1.871728. 
\bibitem{bib:martins_np_2010} S.F. Martins \emph{et al.}, \emph{Nat. Physics}~\textbf{6} (2010) 311. http://dx.doi.org/10.1038/nphys1538
\bibitem{bib:gordienko_pop_2004} S. Gordienko and A. Pukhov, \emph{Phys. Plasmas}~\textbf{12}(4) (2005) 043109. http://dx.doi.org/10.1063/1.1884126
\bibitem{bib:pak_prl_2010} A. Pak \emph{et al.},  \emph{Phys. Rev. Lett.}~\textbf{104}(2) (2010) 025003. http://dx.doi.org/10.1103/PhysRevLett.104.025003
\bibitem{bib:vieira_ppcf_2012} J. Vieira \emph{et al.}, \emph{Plasma Phys. Control. Fusion}~\textbf{54}(12) (2012) 124044. http://dx.doi.org/10.1088/0741-3335/54/12/124044
\bibitem{bib:vieira_prl_2011} J. Vieira \emph{et al.}, \emph{Phys. Rev. Lett.}~\textbf{106}(22) (2011) 225001. http://dx.doi.org/10.1103/PhysRevLett.106.225001
\bibitem{bib:kalmykov_prl_2009} S. Kalmykov \emph{et al.}, \emph{Phys. Rev. Lett.}~\textbf{103}(13) (2009) 135004. http://dx.doi.org/10.1103/PhysRevLett.103.135004
\bibitem{bib:clayton_prl_2002} C.E. Clayton \emph{et al.}, \emph{Phys. Rev. Lett.}~\textbf{88}(15) (2002) 154801. http://dx.doi.org/10.1103/PhysRevLett.88.154801
\bibitem{bib:tzoufras_prl_2008} M. Tzoufras \emph{et al.}, \emph{Phys. Rev. Lett.}~\textbf{101}(14) (2008) 145002. http://dx.doi.org/10.1103/PhysRevLett.101.145002
\bibitem{bib:tsung_prl_2004} F.S. Tsung \emph{et al.}, \emph{Phys. Rev. Lett.}~\textbf{93}(18) (2004) 185002. http://dx.doi.org/10.1103/PhysRevLett.93.185002
\bibitem{bib:allen_pra_1992} L. Allen \emph{et al.}, Phys. Rev. A \textbf{45}, 8185 (1992). 
\bibitem{bib:padgett_pt_2004} M. Padgett \emph{et al.}, Phys. Today \textbf{57}, 5, 35 (2004).
\bibitem{bib:hell_ol_1994} S. W. Hell \emph{et al.}, Opt Lett. \textbf{19}, 780-2 (1994).
\bibitem{bib:wang_natphoton_2012} J. Wang \emph{et al.}, Nature Photonics \textbf{6}, 488?496 (2012).
\bibitem{bib:zhang_science_2020} Z. Zhang \emph{et al.}, Science \textbf{368}, 760-763 (2020). 
\bibitem{bib:tamburini_natphys_2011} F. Tamurini \emph{et al.}, Nat. Physics \textbf{7}, 195?197(2011).
\bibitem{bib:vieira_natcomms_2016} J. Vieira \emph{et al.}, Nat. Comms. \textbf{7}, 10371 (2016).
\bibitem{bib:vieira_prl_2016} J. Vieira \emph{et al.}, Phys. Rev. Lett. \textbf{117}, 265001 (2016).
\bibitem{bib:lehmann_prl_2016} G. Lehmann \emph{et al.} Phys. Rev. Lett. \textbf{116}, 225002 (2016).
\bibitem{bib:lehmann_pre_2019} G. Lehmann \emph{et al.} Phys. Rev. E \textbf{100}, 033205 (2019).
\bibitem{bib:leblanc_nphys_2017} A. Leblanc \emph{et al.} Nat. Phys.13, 440 (2017).
\bibitem{bib:shi_prl_2014} Y. Shi \emph{et al.} Phys. Rev. Lett. \textbf{112}, 235001 (2014).
\bibitem{bib:longman_ol_2020} A. Longman \emph{et al.}, Optics Lett. \textbf{45}, 2187 (2020).
\bibitem{bib:mendonca_pop_2014} J.T. Mendon\c ca \emph{et al.}, Phys. Plasmas \textbf{21}, 033107 (2014).
\bibitem{bib:vieira_prl_2014} J. Vieira and J.T. Mendon\c ca, Phys. Rev. Lett.~\textbf{112}, 215001 (2014).
\bibitem{bib:yu_pop_2014} L.-L. Yu \emph{et al.}, Phys. Plasmas~\textbf{21}, 120702 (2014). 
\bibitem{bib:jain_arxiv_2014} N. Jain \emph{et al.}, arXiv:1410.8762 (2014).
\bibitem{bib:zhang_jap_2016} G.B. Zhang \emph{et al.}, J. Appl. Phys. \textbf{119}, 103101 (2016).
\bibitem{bib:zhang_pop_2016} G.B. Zhang \emph{et al.}, Phys. Plasmas \textbf{23}, 033114 (2016).
\bibitem{bib:ju_njp_2018} L.B. Ju \emph{et al.}, New J. Physics \textbf{20}, 063004 (2018).
\bibitem{bib:blackmann_pre_2019} D. R. Blackman \emph{et al.}, Phys. Rev. E \textbf{100}, 013204 (2019).
\bibitem{bib:pariente_ol_2015} G. Pariente \emph{et al.}, Optics Lett. \textbf{40}, 2037 (2015).
\bibitem{bib:vieira_prl_2018} J. Vieira \emph{et al.}, Phys. Rev. Lett. \textbf{121}, 054801 (2018).
\bibitem{bib:vieira_2020} J. Vieira \emph{et al.}, submitted (2020); A. Gover private communication (2020).
\bibitem{bib:harris_nphys_2015} J. Harris \emph{et al.}, Nat. Phys. \textbf{11}, 629 (2015).
\bibitem{bib:marie_optica_2017} A. Sainte-Marie \emph{et al.}, Optica \textbf{4}, 1298 (2017). 
\bibitem{bib:froula_nphoton_2018} D. Froula \emph{et al.}, Nat. Photonics \textbf{12}, 262 (2018).
\bibitem{bib:palastro_prl_2020} J. P. Palatro \emph{et al.}, Phys. Rev. Lett. 124, 134802 (2020).
\bibitem{bib:debus_prx_2019} A. Debus \emph{et al.} Phys. Rev. X \textbf{9}, 031044 (2019).
\end{thebibliography}
\end{document}